\documentclass[journal,transmag]{IEEEtran}

\usepackage{cite}
\usepackage{amsmath}
\usepackage{amsfonts}
\usepackage{amssymb}
\usepackage{graphicx}
\usepackage{microtype}
\usepackage{booktabs}

\graphicspath{{images/}}

\usepackage[svgnames]{xcolor}
\usepackage[pdftex]{hyperref}
\hypersetup{colorlinks=true, linkcolor=blue, citecolor=ForestGreen, urlcolor=DarkOrchid}

\newcommand{\function}[1]{f_{\mathrm{#1}}}
\newcommand{\weight}[1]{\lambda_{\mathrm{#1}}}
\newcommand{\BB}{{\{0, 1\}}}
\newcommand{\Nf}{N_{\mathrm{f}}}
\newcommand{\Nc}{N_{\mathrm{c}}}

\begin{document}

\title{Quantum Annealing Applied to De-Conflicting Optimal Trajectories for
Air Traffic Management}

\author{\IEEEauthorblockN{
Tobias Stollenwerk\IEEEauthorrefmark{1},
Bryan O'Gorman\IEEEauthorrefmark{2,3,4},
Davide Venturelli\IEEEauthorrefmark{3,5},
Salvatore Mandr{\`a}\IEEEauthorrefmark{3,4},
Olga Rodionova\IEEEauthorrefmark{6},\\
Hok K. Ng\IEEEauthorrefmark{6},
Banavar Sridhar\IEEEauthorrefmark{6},
Eleanor G. Rieffel\IEEEauthorrefmark{3}, and
Rupak Biswas\IEEEauthorrefmark{3},
}
\IEEEauthorblockA{\IEEEauthorrefmark{1}German Aerospace Center (DLR), Cologne, Germany 51147}
\IEEEauthorblockA{\IEEEauthorrefmark{2}University of California, Berkeley, CA 94720}
\IEEEauthorblockA{\IEEEauthorrefmark{3}Quantum Artificial Intelligence Laboratory (QuAIL), NASA Ames Research Center, Moffett Field, CA 94035}
\IEEEauthorblockA{\IEEEauthorrefmark{4}Stinger Ghaffarian Technologies Inc., Greenbelt, MD 20770}
\IEEEauthorblockA{\IEEEauthorrefmark{5}Research Institute for Advanced Computer Science, Universities Space Research Association (USRA), Mountain View, CA 94035}
\IEEEauthorblockA{\IEEEauthorrefmark{6}NASA Ames Research Center, Moffett Field, CA 94035}
}

\IEEEtitleabstractindextext{
\begin{abstract}
We present the mapping of a class of simplified air traffic management (ATM)
problems (strategic conflict resolution) to quadratic unconstrained boolean
optimization (QUBO) problems.  The mapping is performed through an original
representation of the conflict-resolution problem in terms of a conflict graph,
where nodes of the graph represent flights and edges represent a potential
conflict between flights.  The representation allows a natural decomposition of
a real world instance related to wind-optimal trajectories over the Atlantic ocean
into smaller subproblems, that can be discretized and are
amenable to be programmed in quantum annealers.  In the study, we tested the new
programming techniques and we benchmark the hardness of the instances
using both classical solvers and the D-Wave 2X and D-Wave 2000Q quantum chip.
The preliminary results show that for reasonable modeling choices the most challenging subproblems which are
programmable in the current devices are solved to optimality with 99\% of probability within a second of annealing time.
\end{abstract}

}

\maketitle

%%%%% INTRODUCTION %%%%%%
\section{Introduction}
One of the main functions of Air Traffic Control (ATC) is ensuring safe flight
progress in the shared airspace. This in particular involves resolving potential
conflicts between flights, where “a conflict” stands for a violation of
separation norms established in the airspace.

There is an overall increase in air traffic over the last decades and this trend
is believed to continue.  As a result, ATC workload is constantly increasing.
Nowadays the flights are typically assigned the predefined routes from the air
traffic network, which is becoming saturated.  With the limited airspace
available, novel approaches are necessary to meet the increasing air
traffic demand in the coming decades.  
One promising approach that addresses both traffic congestion and fuel costs is to start with wind-optimal trajectories, i.e. the route that each flight would take to minimize fuel costs if there were no other flights~\cite{ng_optimizing_2014}. 
Such wind-optimal trajectories will conflict with each other~\cite{rodionova16} and thus be deconflicted.
Conflict detection and resolution is a complex problem
which has been studied for the decades \cite{kuchar2000review,emami-2012}.

Quantum annealing is a promising computational method which became increasingly
important in recent years.  This development is driven also by first
commercially available quantum annealing device by the company D-Wave Systems.
In addition to studying the fundamental properties of quantum annealing, it is
imperative to find possible real world application for this technology.  Hard
operational planning problems are a promising candidate for the
latter~\cite{Rieffel2015, AAAI148614, Venturelli2015}.

In this work, we investigate the feasibility of applying quantum annealing to
solve the conflict resolution problem for wind-optimal trajectories. To be
amenable to a D-Wave quantum annealer, the conflict-resolution problem has to be
formulated as a quadratic unconstrained binary optimization (QUBO) problem. Given
the discrete nature of QUBO problems, a tunable discretization must be introduced.
Nevertheless, while treating trajectories as continuous functions can be beneficial
\cite{rodionova16}, it is also more computational demanding. On the contrary, a QUBO
formulation of the de-conflitting problem allows to “tune” the discretization to trade
between quality of solutions and computational effort.
For the main part of the paper, we restrict ourselves to a
simplified version of the problem by considering departure delays only while
neglecting maneuvers.
We present a detailed study of the structure of this problem which provides
insights beyond the scope of quantum annealing.

In particular, we perform the following analyses

\begin{itemize}
\item
Given the wind-optimal trajectories, we extract natural subsets of the overall
problem and study their hardness.  We found that the problems are hard in
general and become harder as we increase the maximum allowed value for the
departure delays.

\item
Restrictions to the configurations space are necessary for the reformulation of
the problem as a QUBO.\@ Therefore, we employ classical solvers to investigating
the influence of discretization on the solution quality.  As a result, we found
that finer discretization increases the solution quality and sufficiently large
maximum allowed value for the departure delays is enough for an acceptable solution

\item
We demonstrate the mapping of the deconflicting problem to a QUBO formulation
for models with (see Appendix) and without maneuvers (in main text).  In the
course of this, we investigate the sufficient penalty weights for the hard
constraints in the problem.  Here we found that these penalty weights are
largely independent of the problem instances.

\item
We investigate the possibility to embed deconfliction-derived QUBO instances
onto the D-Wave quantum annealer. More precisely, we were able to embed and
run smaller problem instances and found that finer model discretizations as
well as larger problem sizes decrease the success probability due to the limited
precision of the D-Wave 2X machine.

\end{itemize}

The paper is organized as follows: We begin by formulating the
conflict-resolution problem as a combinatorial optimization problem and
describing the preprocessing necessary for this mapping in
Section~\ref{sec:problem_specification}.  In Section~\ref{sec:instances} we
investigate the structure and hardness of problem instances before we study the
impact of discretization on the solution quality in
Section~\ref{sec:discretization}.  Afterwards, we discuss the mapping of the
problem to a QUBO in Section~\ref{sec:mapping}.  We report on the
embeddability of the QUBO instances and their solution quality on a D-Wave 2X
device in Section~\ref{sec:qa}.  Finally, we conclude with discussion on improvements and future works.
In the Appendix, we present more general
mappings (including maneuvers) of the original deconflicting problem to QUBOs.

%%%%%% PROBLEM SPECIFICATION %%%%%%
\section{Problem specification}
\label{sec:problem_specification}

The basic input of the conflict-resolution problem is a set of optimal flight
trajectories (space-time paths). Such trajectories are the results of
optimizations performed by the flight operators.

In the present study, we consider wind-optimal trajectories. Such trajectories
are obtained by minimizing the fuel cost over the routes with given origins and
destinations and desired departure times in the presence of forecast winds.
Because of the correlation between such trajectories arising from exploiting
favorable winds, these trajectories are likely to conflict; that is, two or more
aircraft are likely to get dangerously close to each other if their optimal
trajectories are followed without modification.  The goal thus is to modify the
trajectories to avoid such conflicts.

In theory, the configuration space consists of all physically realistic
trajectories; in practice, computational limits constrain us to consider certain
perturbations of the optimal trajectories. The simplest way to perturb a
trajectory is to delay the corresponding flight on the ground prior to
departure. These are the type of perturbations we mainly analyze in this work.
We also consider local spatial modifications of the trajectories so that no new
potential conflicts are induced. Such local modifications can then parametrized
as effective additional delay.  Previous work~\cite{rodionova16} additionally
considered a global modification of the trajectory geometrical shape.

A full accounting of the cost of such modifications would include the
cost of departure delays, the change in fuel cost due to perturbing the
trajectories, the relative importance of each flight, and many other factors.
As in previous work, we consider only the total, unweighted arrival delay,
aggregated equally over all of the flights.

Formally, each optimal trajectory $\mathbf x_i = {\left(x_{i,
t}\right)}_{t=\tau_{i,0}}^{\tau_{i,1}}$ is specified as a time-discretized path
from the departure point $x_{i, \tau_{i,0}}$ at time $\tau_{i,0}$ to the
arrival point $x_{i, \tau_{i,1}}$ at time $\tau_{i,1}$.  For each flight $i$,
the geographical coordinates $x_{i,t}$ (as latitude, longitude, and altitude)
are specified at every unit of time (i.e.\ one minute) between $\tau_{i,0}$ and
$\tau_{i,1}$.

For notational simplicity, suppose momentarily that each trajectory $\mathbf
x_i$ is modified only by introducing delays between time steps.  Let
$\delta_{i, t}$ be the accumulated delay of flight $i$ at the time that it
reaches the point $x_{i, t}$, and let $\delta^*_{i, t}$ be the maximum such
delay at the point (given the modifications under consideration). Then, the
total delay over $N_f$ flights is
\begin{equation}
  D = \sum_{i=1}^{\Nf} \delta_{i, \tau_{i,1}}.
\end{equation}

\begin{figure}[htpb]
    \begin{center}
        \includegraphics[width=0.45\textwidth]{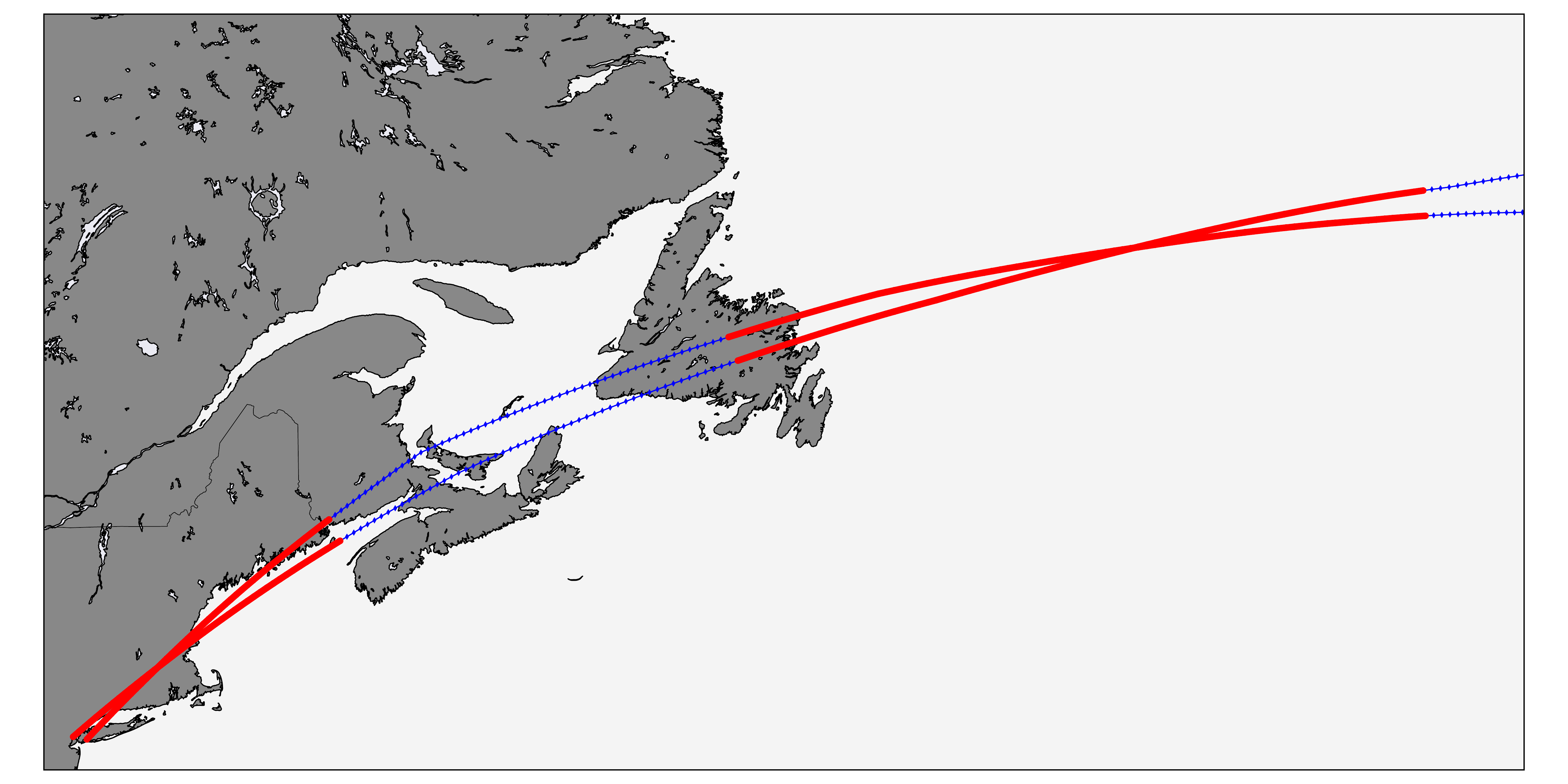}
    \end{center}
    \begin{center}
        \includegraphics[width=0.35\textwidth]{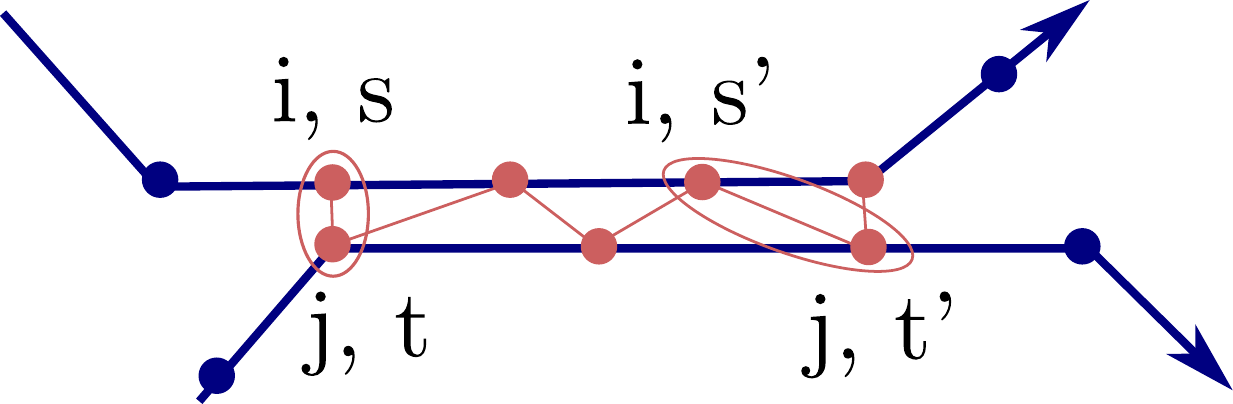}
    \end{center}
    \caption[Conflict example]{Top:Example of two
    potential conflicts between a pair of transatlantic flights originating on
    the East Coast of the USA.
    Bottom: Conflict definition for a pair of flights $i$ and $j$. If we have two pairs of conflicting trajectory points $\{(i, s), (j, t)\}, \{(i, s'), (j, t')\}$, all the intermediate points must also be in conflict to each other in order to qualify as a conflict cluster.
    } \label{fig:example_parallel_conflict}
\end{figure}

A pair of flights $(i, j)$ are in spatial conflict with each other if any pair
of points from their trajectories is in conflict. That is, a pair of
trajectory points $(x_{i,s}, x_{j,t})$, at time $s$ and $t$ respectively, 
conflict if their spatial and temporal
separations are both less than the respective mandatory separation standards
$\Delta_x$ and $\Delta_t$:
\begin{subequations}
\begin{equation}
\left\|x_{i, s} - x_{j,t}\right\| < \Delta_x,
\end{equation}
and
\begin{equation}
\left|\left(s + \delta_{i, s}\right) - \left(t + \delta_{j,t}\right)\right|
< \Delta_t.
\end{equation}
\end{subequations}
For the North Atlantic oceanic airspace, the separation standard are set to be:
30 nautical miles for horizontal separation $\Delta_x$ and 3 minutes for
temporal separation $\Delta_t$.
Observe that the latter condition can be met for some $\left(\delta_{i,s},
\delta_{j,t}
\right) \in [0, \delta^*_{i,s}] \times [0, \delta^*_{j,t}]$ if and only if
\begin{equation}
\max \left\{\delta^*_{i, s}, \delta^*_{j, t}\right\} + \Delta_t > |s - t|,
\end{equation}
in which case we call the pair of trajectory points \emph{potentially}
conflicting. Let us partition the set $C$ of potentially conflicting pairs of trajectory points
into disjoint sets, or clusters, $C_k$:
\begin{equation}
C =  \bigcup_{k} C_k,
\end{equation}
such that if $\left\{(i, s), (j, t)\right\}, \left\{(i', s'), (j', t')\right\}
\in C_k$ for some $k$ then $i = i' < j = j'$ and for all $s'' \in [\min \{s,
s'\}, \max\{s, s'\}]$ there exists some ${t'' \in [\min \{t, t'\}, \max\{t,
t'\}]}$ such that ${\left\{(i, s''), (j, t'')\right\} \in C_k}$ and vice versa.
We will further refer to such clusters $C_k$ simply as the conflicts.
Note that choosing $i < j$ in the definition is just a convention to uniquely determine the two flights involved in the conflict, and the index $k$ has no meaning other than uniquely identifying conflicts.
The purpose of clustering in this way is to extract a single constraint for each cluster than address all of the potential conflicts therein.
Figure \ref{fig:example_parallel_conflict} shows an example of two such conflict clusters.
Thus every
conflict $k$ is associated with a pair of flights $I_k = \{i, j\}$.  Let $K_i =
\left\{k \middle| i \in I_k \right\}$ be the set of conflicts to which flight
$i$ is associated, $\Nc$ the number of conflicts.

Having identified disjoint sets of conflicts, we relax the supposition that the
trajectory modifications only introduce delays between time steps.  Instead, we
consider modifications to the trajectories that introduce delays local to
particular conflicts.  Specifically, the configuration space consists of the
departure delays $\mathbf d = {\left(d_i\right)}_{i=1}^{\Nf}$ and the set of
local maneuvers $\mathbf a_{\mathbf k} = {\left(\mathbf a_k\right)}_k$, where
$\mathbf a_k$ represents some parameterization of the local maneuvers used to
avoid conflict $k$.  Let $d_{i, k} (\mathbf d, \mathbf a_{\mathbf k})$ be the
delay introduced to flight $i$ at conflict $k$, as a function of the departure
delays and local maneuvers.  With this notation, we can write the total delay
as
\begin{equation}
D =
\sum_{i = 1}^{\Nf}
\left(d_i + \sum_{k \in K_i} d_{i, k}\right).
    \label{eq:total-delay}
\end{equation}
This is the quantity we wish to minimize subject to avoiding all potential conflicts.

We focus on the case where conflicts will be avoided only by the introduction of extra delays, leaving for future work the introduction of maneuvering choices.

Let
\begin{equation}
\label{eq:accum-delay}
D_{i, k} = d_i + \sum_{k' \in K_{i, k}} d_{i,k'}
\end{equation}
be the accumulated delay of flight $i$ by the time it reaches conflict $k$,
where $K_{i, k} = \left\{k' \in K_i \middle| k' < k\right\}$. We assume that
the set of conflicts $K_i$ associated with flight $i$ is indexed in temporal
order, i.e.\ if $k' < k$ and $k, k' \in K_i$, then flight $i$ reaches conflict
$k'$ before conflict $k$. For simplicity, we assume that no delay is introduced
during a conflict, so that $\delta_{i, s} = D_{i, k}$ for all $s$ associated
with flight $i$ in conflict $k$. The pairs of conflicting trajectory points
associated with conflict $k$ are given by
\begin{equation}
T_k =
\left\{
(s, t) \middle| \{(i, s), (j, t)\} \in C_k, i < j
\right\}.
\end{equation}
Thus the potential conflict is avoided only if
\begin{equation}
\label{eq:accum-delay-diff}
D_k = D_{i, k} - D_{j, k}
\notin
B_k
\end{equation}
where
\begin{align}
B_k &=
\bigcup_{(s, t) \in T_k}
\left(-\Delta_t + t - s, \Delta_t + t - s\right)
=
[\Delta^{\min}_k, \Delta^{\max}_k], \\
\Delta^{\min}_k &= 1 - \Delta_t + \min_{(s, t) \in T_k} \{t - s\},\\
\Delta^{\max}_k &= \Delta_t - 1 + \max_{(s, t) \in T_k} \{t - s\}.
\end{align}

In the remainder of this paper, we focus on the simplified problem in which
only departure delays are allowed.  In this case, the configuration
space is simply $\mathbf d = {\left(d_i\right)}_{i=1}^{\Nf}$, the cost function
(Eq.~\ref{eq:total-delay}) transforms into $D = \sum_{i = 1}^{\Nf} d_i$, and the
constraints become $d_i - d_j \notin B_k$ for all $k$.

%%%%%% INSTANCES %%%%%%
\section{Instances}\label{sec:instances}
We test
on realistic
instances of the problem, using the precalculated wind-optimal trajectories
for transatlantic flights on July 29, 2012~\cite{rodionova16}.
This data consists of 984 flights each of which has a constant cruising altitude and constant speed.
However, our
methods can be generalized to instances without these special properties.

\begin{figure}[htpb]
\includegraphics[width=\columnwidth]{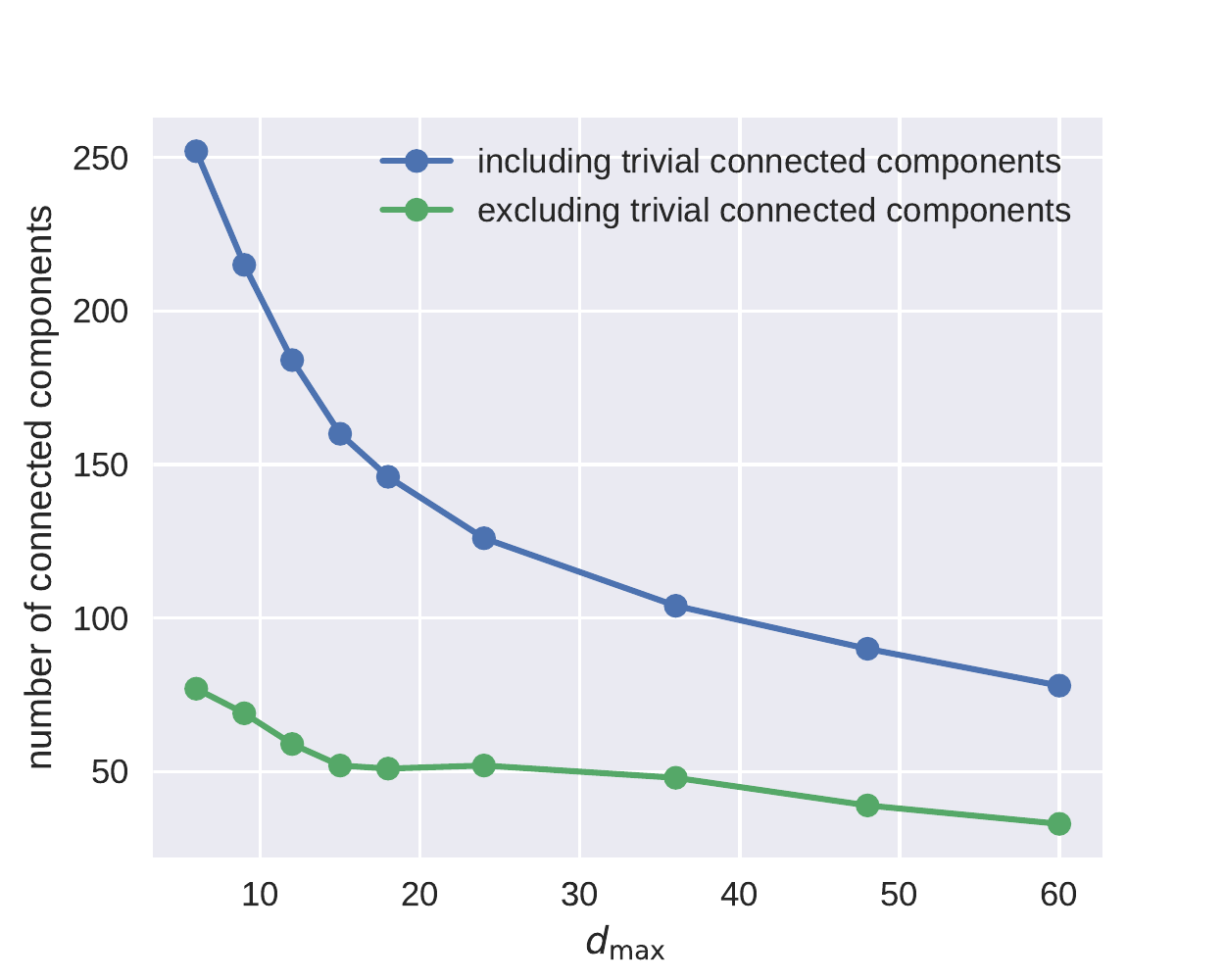}
\caption[Number of non-trivial connected components vs. $d_{\max}$]{Number of connected components versus $d_{\max}$.}
\label{fig:num-CCs-vs-dmax}
\end{figure}

\begin{figure}[htpb]
\includegraphics[width=\columnwidth]{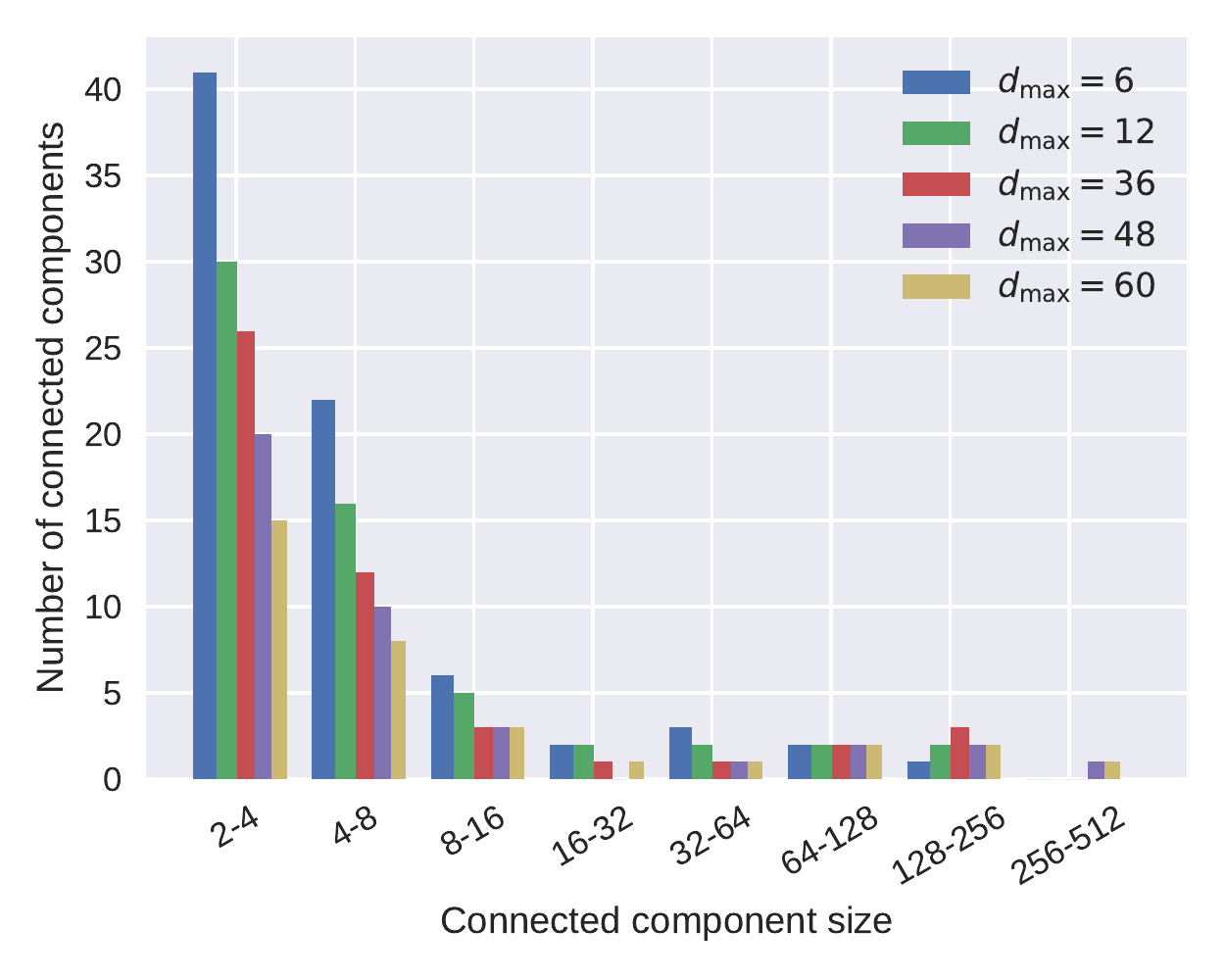}
\caption[Histogram of connected component sizes]{Histogram of the connected component size for various values of the maximum delay time $d_\text{max}$.}
\label{fig:hist-CC-sizes}
\end{figure}

To identify the instances of the conflict-resolution problem we construct a
\emph{conflict graph}, whose vertices correspond to flights and which has an
edge between a pair of vertices if there is at least one potential conflict
between the corresponding flights. Note that the conflict graph for a given set
of trajectories depends on the parameters of the problem.  In the case of only
departure delays, whether or not a potential conflict, and thus an edge in the
conflict graph, exists between two flights is a function of the maximum
allowable departure delay $d_{\max}$.
For
a certain value of $d_{\max}$, the conflict graph may contain several connected
components, which can be considered as smaller, independent instances.
Figure~\ref{fig:num-CCs-vs-dmax} shows this dependence of the number of
connected components (both including and excluding trivial connected
components, i.e.\ those containing a single vertex) on the maximum delay
$d_{\max}$, and Figure~\ref{fig:hist-CC-sizes} shows the distribution of the
sizes of the connected components for various values of $d_{\max}$.  As
$d_{\max}$ increases, the conflict graph becomes denser; at some point, the
conflict graph saturates (though not necessarily as the complete graph), with
every spatial conflict indicating a potential conflict.  Interestingly, most of
the connected components are very small; for example, with $d_{\max}= 60$
minutes, approximately $75\%$ of the connected components contain no more than
$10$ flights.

In the remainder of this paper, we consider sets of smaller instances
corresponding to the connected components of the conflict graph from the larger
single instance for various values of $d_{\max}$, for the given flight set. Let
$\mathcal I_{d_{\max}}$ be the set of such instances for a particular value of
$d_{\max}$, excluding trivial instances.  We say that an instance is trivial if
there are no conflicts when all flights therein depart without delay; in
particular, this includes instances containing only a single flight.

\begin{figure}[htpb]
\includegraphics[width=0.95\columnwidth]{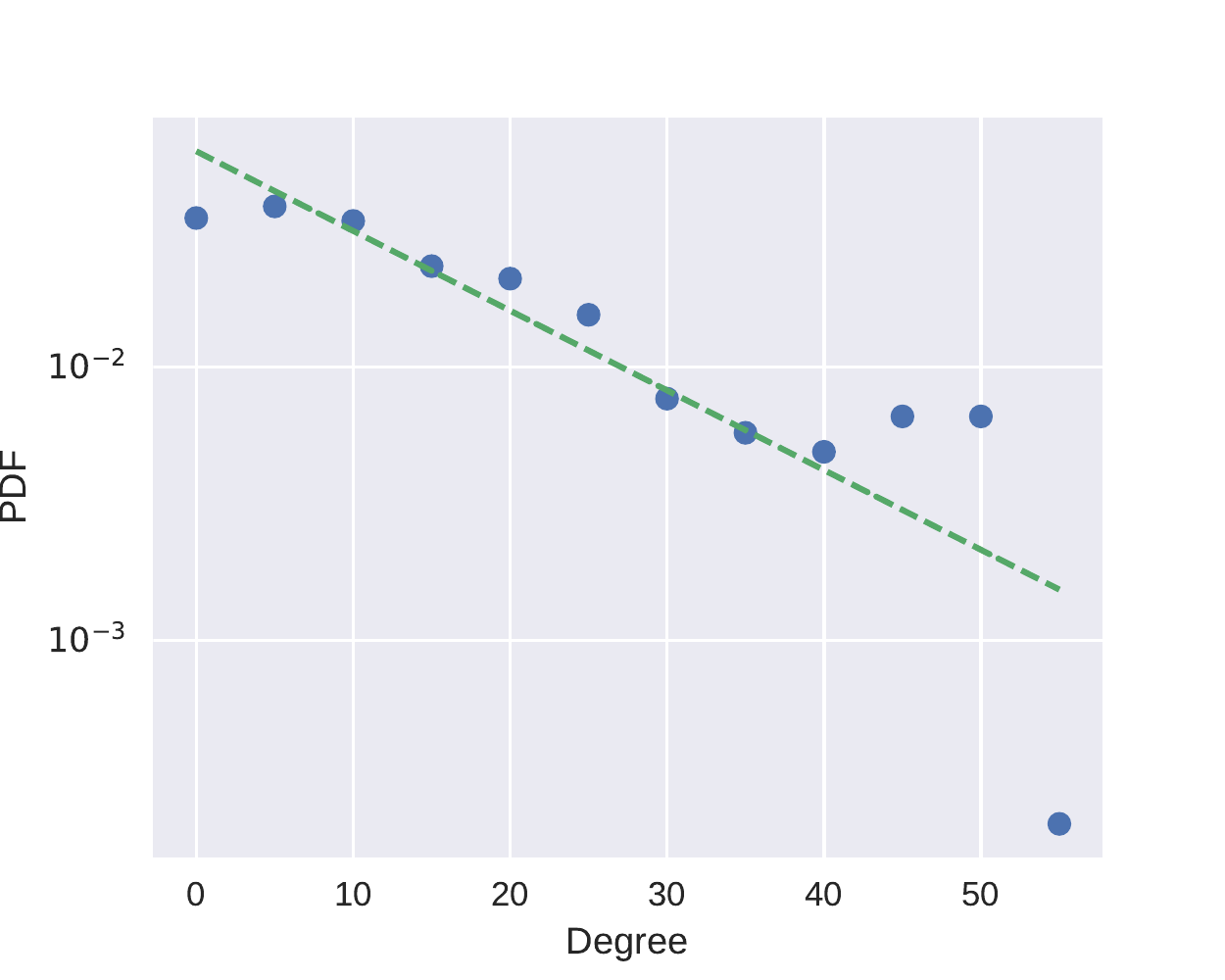}
\caption[Histogram of degrees]{Histogram of the degrees of vertices in the conflict graph for $d_{\max} = 60$.
The distribution of the degrees approximately follows a power law, with the exponent depending on $d_{\max}$.
}
\label{fig:hist-degree}
\end{figure}

\begin{figure}[htpb]
\includegraphics[width=0.95\columnwidth]{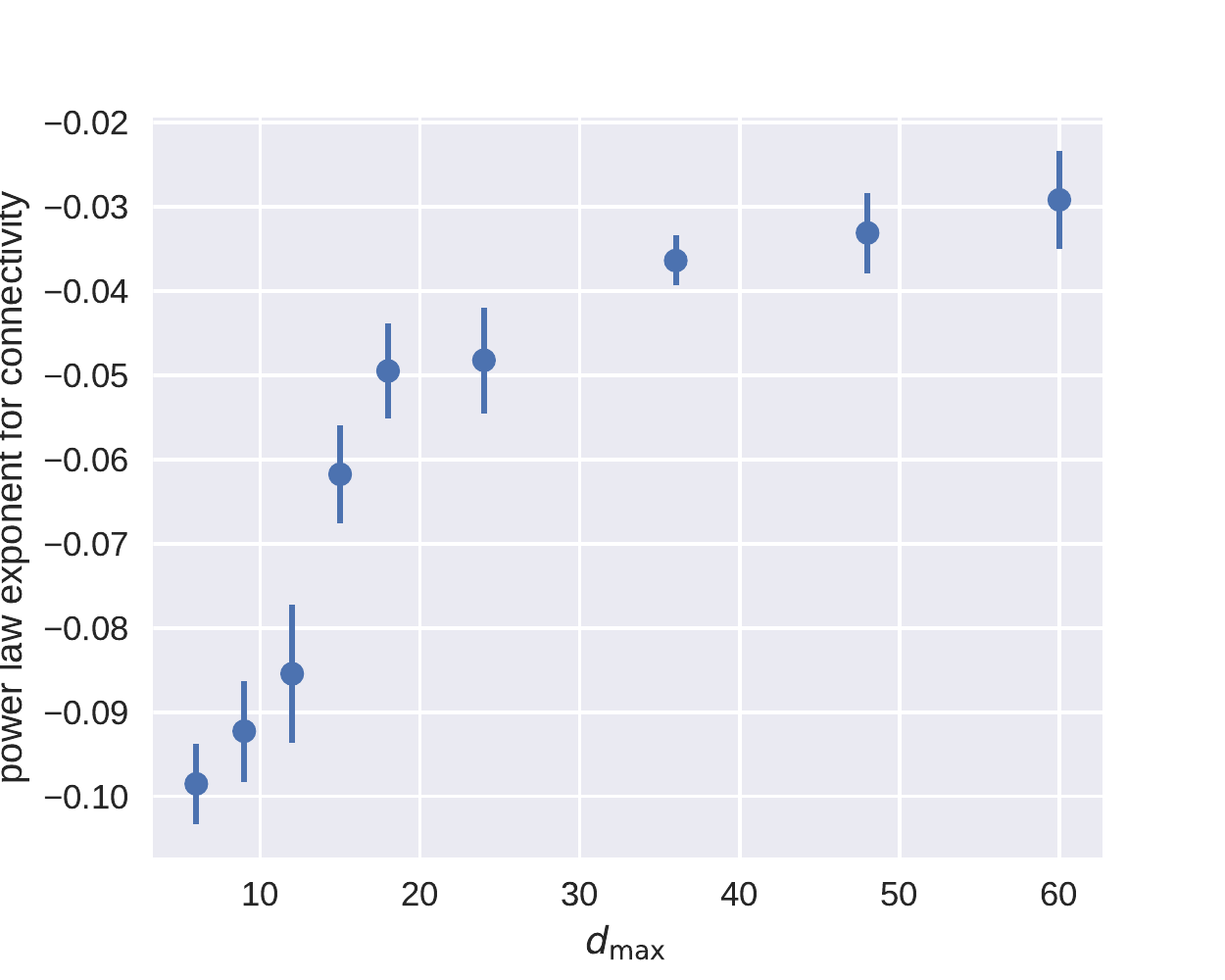}
    \caption[Power-law exponent vs. $d_{\max}$]{Empirical power-law exponent versus $d_{\max}$. The error bars indicate the error obtained from the linear regression.}
\label{fig:exponent-vs-dmax}
\end{figure}

As a part of our analysis, we also studied the probability distribution of the
degree of vertices in the conflict graph. In other words, the number of flights
for which a given flight share a potential conflict with.
Figure~\ref{fig:hist-degree} shows the distribution of degrees of vertices in
the conflict graph for $d_{\max}=60$, which seem to approximately coincide with
a power law, i.e.  the number of vertices with degree $d$ is proportional to
$d^{\alpha}$.  This is consistent with a so-called ``small-world'' model
believed to be typical of many real-world graphs~\cite{barabasi:99}, which are
generated by preferential attachment and resultingly contain a few number of
highly-connected hubs, as is the case with air traffic.
Figure~\ref{fig:exponent-vs-dmax} shows the dependence of this
empirical power-law exponent $\alpha$ as a function of $d_{\max}$.  As
$d_{\max}$ increases, the exponent decreases.  The larger the delay, the less
the structure of the trajectories matters and the flatter the distribution of
degrees in the conflict graph.

In many cases, generally hard problems are easy when restricted to tree-like
instances~\cite{bertele1972, halin1976s}.  For example, if the conflict graph
here is a tree, then the optimum could be easily found by propagating the
delays along the tree; on the other hand, if the conflict graph is a complete
graph, finding the optimum is much harder.  The tree-width of a graph
formalizes this notion of tree-likeness, ranging from $1$ for a tree to $n-1$
for fully connected graph.  We examine the treewidth of the connected
components as a proxy for the hardness of the instances they
represent.

\begin{figure}[htpb]
 \includegraphics[width=\columnwidth]{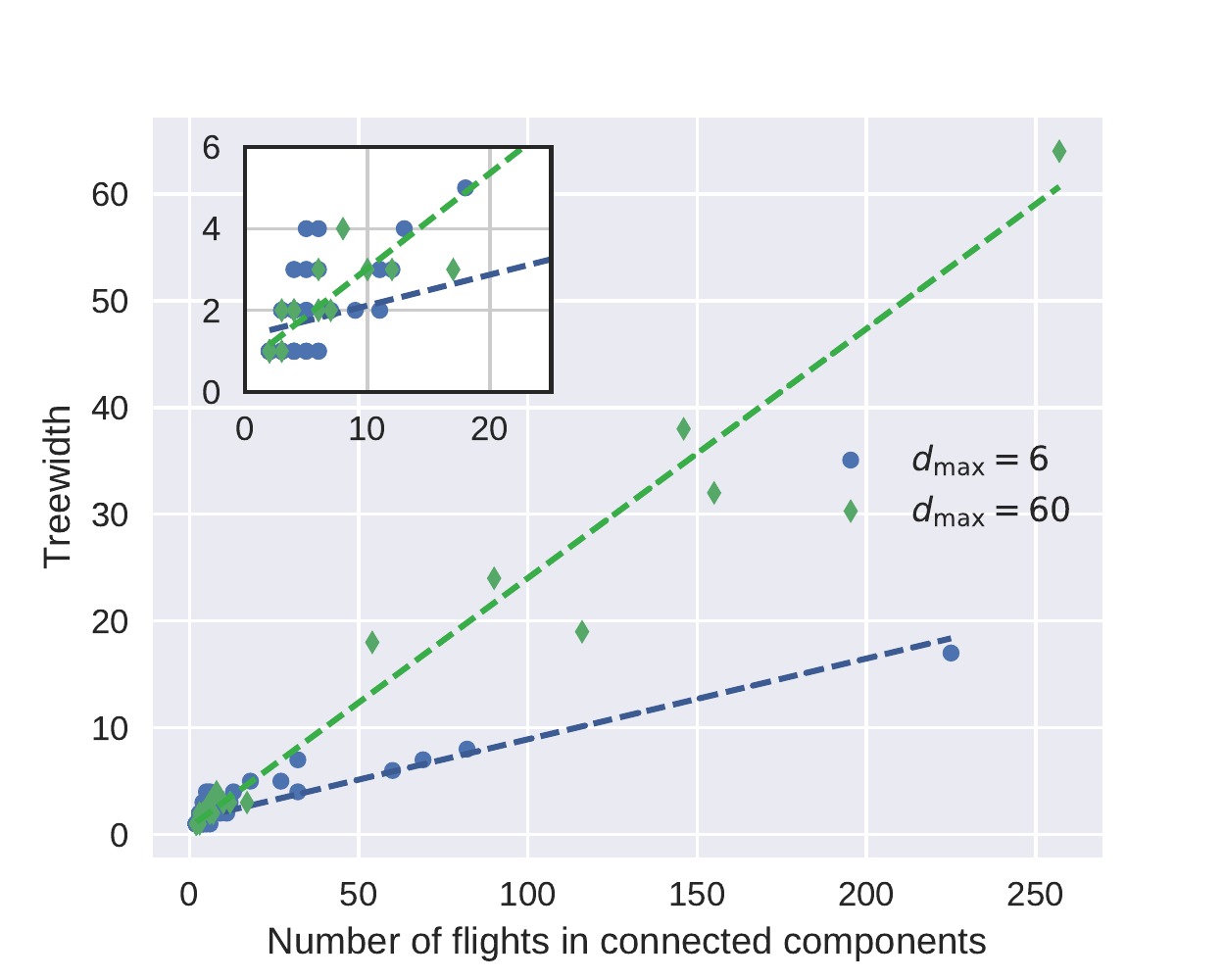}
 \caption[Correlation between connected component size and treewidth]{
 The treewidths of connected components versus their sizes for various values of $d_{\max}$.
 The correlation is approximately linear, with a slope $\gamma$ that depends on $d_{\max}$.
 The linear fit is representing the trend in the region with number of flights greater than $50$.
 }
 \label{fig:tw-vs-CC-size}
 \end{figure}

\begin{figure}[htpb]
 \includegraphics[width=\columnwidth]{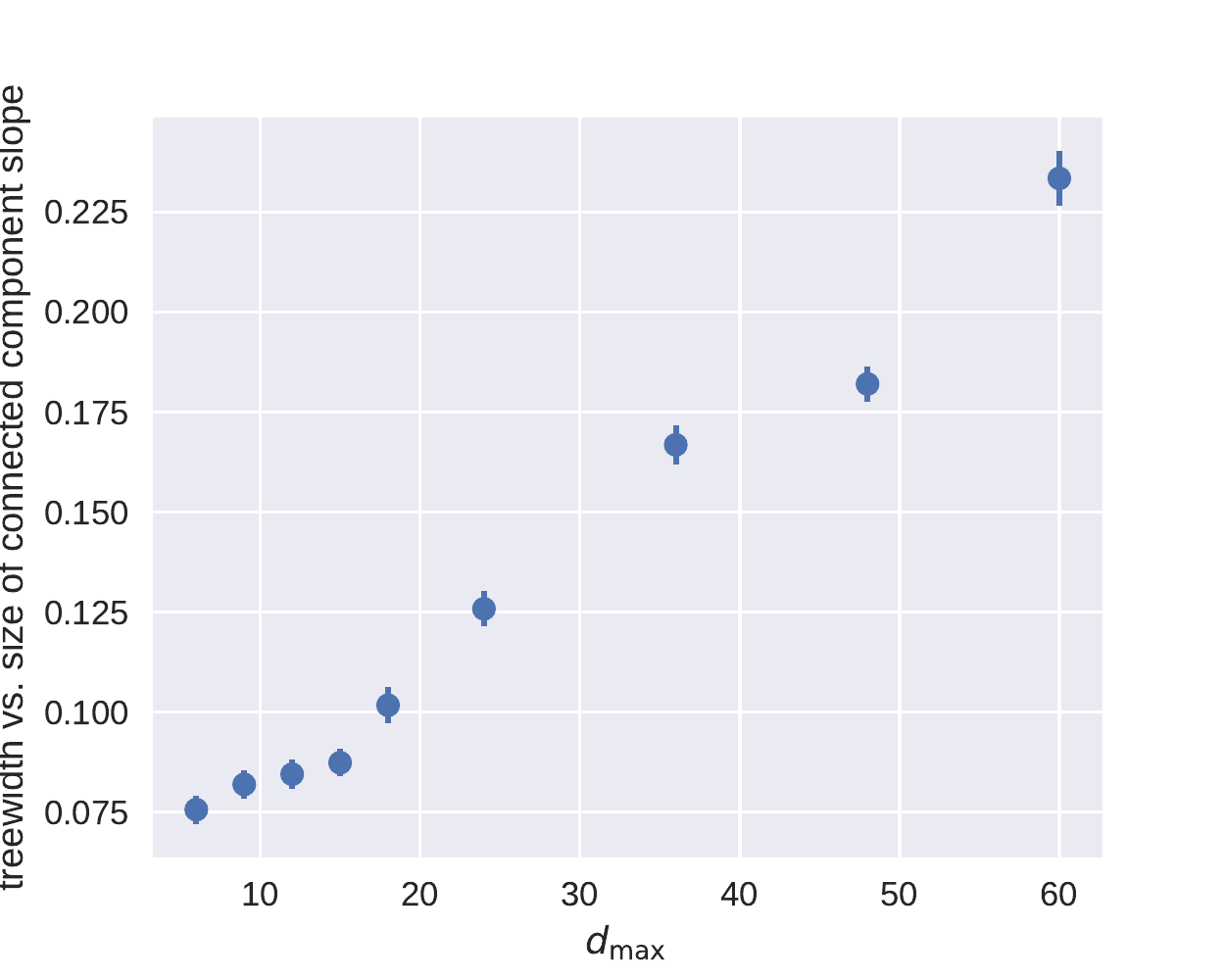}
 \caption[Treewidth-size correlation coefficient vs. $d_{\max}$]{Slope $\gamma$ as a function
     of the maximum delay time. The error bars indicate the error obtained from the linear regression.}
 \label{fig:treewidth-size-correlation}
 \end{figure}

Figure~\ref{fig:tw-vs-CC-size} shows that the treewidth of a connected
component scales approximately linearly with its size.  This suggests that
realistic instances of the deconflicting are indeed hard, and not restricted to
easier (bounded tree-width) instances of the generally hard problem.  Moreover,
the correlation $\gamma$ between the tree-width of a connected component and
its size increases with $d_{\max}$, as shown in
Figure~\ref{fig:treewidth-size-correlation}.  The larger $d_{\max}$, the more
potential conflicts there are; restricting $d_{\max}$ also restricts the number
of conflicts.

%%%%%% DISCRETIZATION %%%%%%
\section{Discretizing the configuration space}
\label{sec:discretization}
To apply quantum annealing to the deconflicting problem, we must encode the
configuration space $\mathbf d$ in binary-valued variables.  To do so, we must
first discretize and bound the allowed values.  Let $\Delta_d$ be the
resolution of the allowed delays and $d_{\max} = N_d \Delta_d$ the maximum
allowed delay, so that $d_i \in \left\{\Delta_d l \middle| l \in [0, 1, \ldots,
N_d]\right\}$, where $d_i$ is the departure delay of flight $i$.  The larger the configuration space is, the more qubits are
needed to encode it, and so determining the effect of this discretization on
solution quality is crucial to the effective use of quantum annealing.  To do
so, we solve the conflict-resolution problem with departure delays only for
various delay resolutions and upper bounds and compare the various optima to
the continuous problem without restrictions (other than non-negativity) on the
delays.

We consider two sets of instances, $\mathcal I_{18}$ and $\mathcal I_{60}$.
For $\mathcal I_{18}$, the exact optima are found by modeling the problem as a
constraint satisfaction problem~\cite{numberjack}; the largest instance in
$\mathcal I_{18}$ has $50$ flights and $104$ potential conflicts.

The instances in $\mathcal I_{60}$ are much larger and harder 
(the largest instance in $\mathcal I_{60}$ has $257$ flights and $4068$ potential conflicts);
we solved them
by mapping to QUBO (as described in the next section) and then using the
Isoenergetic Cluster Method (ICM) (a rejection-free cluster algorithm for spin
glasses that greatly improves thermalization)~\cite{zhu2015}, which has been
shown to be one of the fastest classical heuristic to optimize QUBO
problems~\cite{mandra2016}.  Because ICM is a classical method, the penalty
weights can be set arbitrarily large, ensuring that the desired constraints are
satisfied.
ICM is not guaranteed to return the global optimum in general. However, for
the sizes of instances to which we applied ICM the results are sufficiently
well converged to conclude that the solution found is indeed globally optimal
with exceedingly high probability.

%%%%%%%%%%%%%%%%%%%%%%%%%%%%%%%%%%%%%%%%%%%%%%%%%%%%%%%%%%%%%%%%%%%%%%%%%%%%%%%%
% I_18
%%%%%%%%%%%%%%%%%%%%%%%%%%%%%%%%%%%%%%%%%%%%%%%%%%%%%%%%%%%%%%%%%%%%%%%%%%%%%%%%

\begin{figure}[htpb]
\centering
\includegraphics[width=0.45\textwidth]{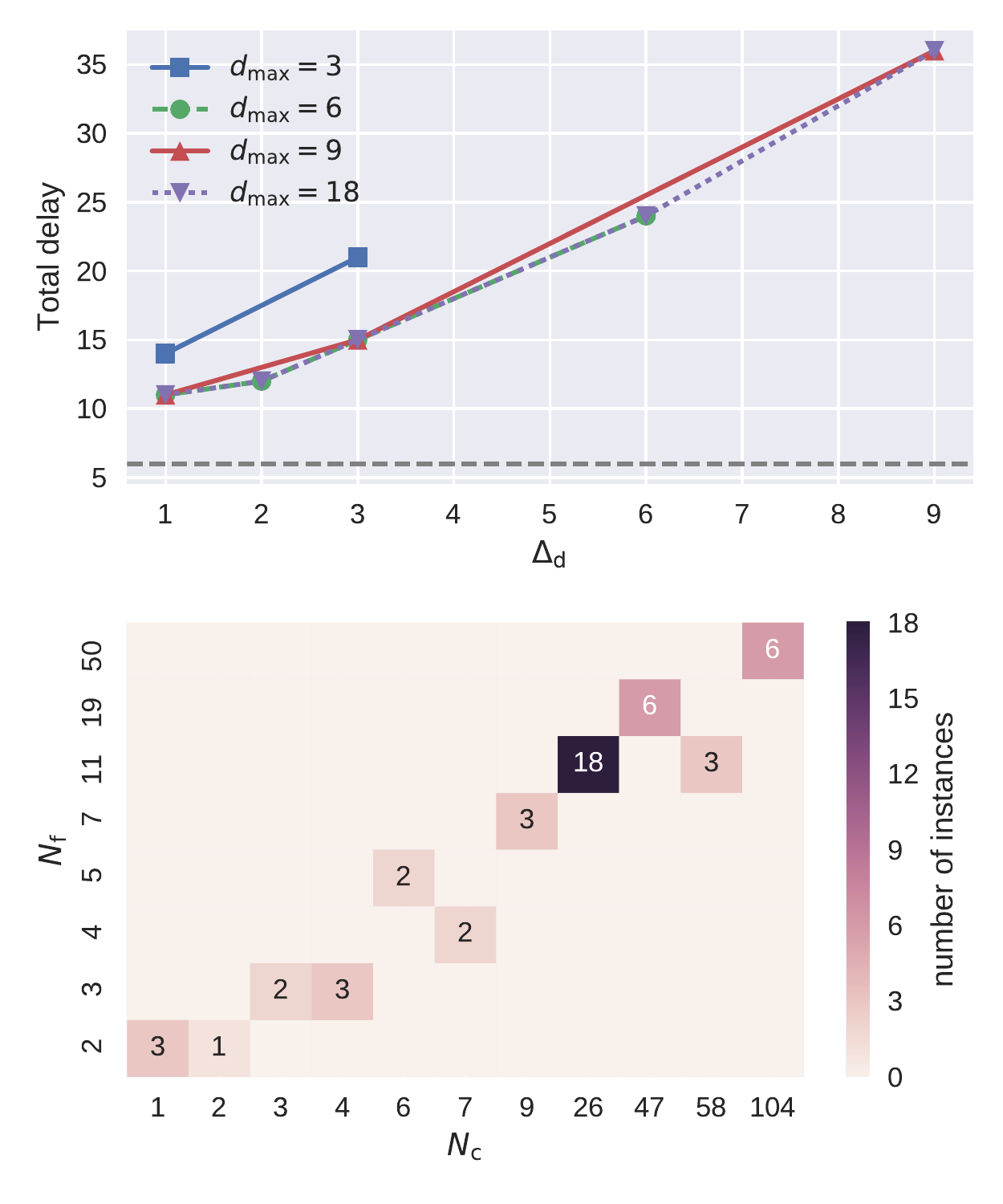}
\caption[Effect of discretization on solution quality]{Influence of discretization on the solution quality. Top: Minimum total delay
of a problem instance from $\mathcal{I}_{18}$ with $19$ flights and $47$
    conflicts for various values of $\Delta_d$ and $d_{\max}$.  Bottom: Results for continuous delay variables with upper bounds $d_\text{max}$.
    We show the minimum upper bound $d_\text{max}$ necessary to obtain same result as that without bounding the delay. We used various instances in $\mathcal I_{18}$.
    The color code shows the number of instances with the same total delay.
}
\label{fig:delay_only_cp_results}
\end{figure}

Figure~\ref{fig:delay_only_cp_results} shows the minimum total delay of a
problem instance with $19$ flights and $47$ potential conflicts from $\mathcal I_{18}$ for various
values of $\Delta_d$ and $d_{\max}$.  With the exception of the small maximum
delay $d_\text{max} = 3$, the total delay of the solutions is nearly
independent of the maximum delay.  The total delay is non-decreasing with
respect to the coarseness $\Delta_d$ of the discretization for a fixed maximum
delay $d_{\max}$, and non-increasing with respect to $d_{\max}$ for a fixed
$\Delta_d$.  Since the original data is discretized in time in units of $1$
minute, $\Delta_d=1$ yield the same result as a continuous variable with the
same upper bound.  Above some threshold value $d^0_\text{max}$, further
increasing the maximum delay does not decrease the minimum total delay.  With
one exception, we found that for all the investigated problem instances
$d^0_\text{max}\leq6$ minutes (see figure~\ref{fig:delay_only_cp_results}).
Therefore we conclude, that a moderate maximum delay is sufficient even for
larger problem instances.  On the other hand, the delay discretization should
be as fine as possible to obtain a high quality solutions.

%%%%%%%%%%%%%%%%%%%%%%%%%%%%%%%%%%%%%%%%%%%%%%%%%%%%%%%%%%%%%%%%%%%%%%%%%%%%%%%%
% I_60
%%%%%%%%%%%%%%%%%%%%%%%%%%%%%%%%%%%%%%%%%%%%%%%%%%%%%%%%%%%%%%%%%%%%%%%%%%%%%%%%

\begin{figure}[htpb]
  \includegraphics[width=\columnwidth]{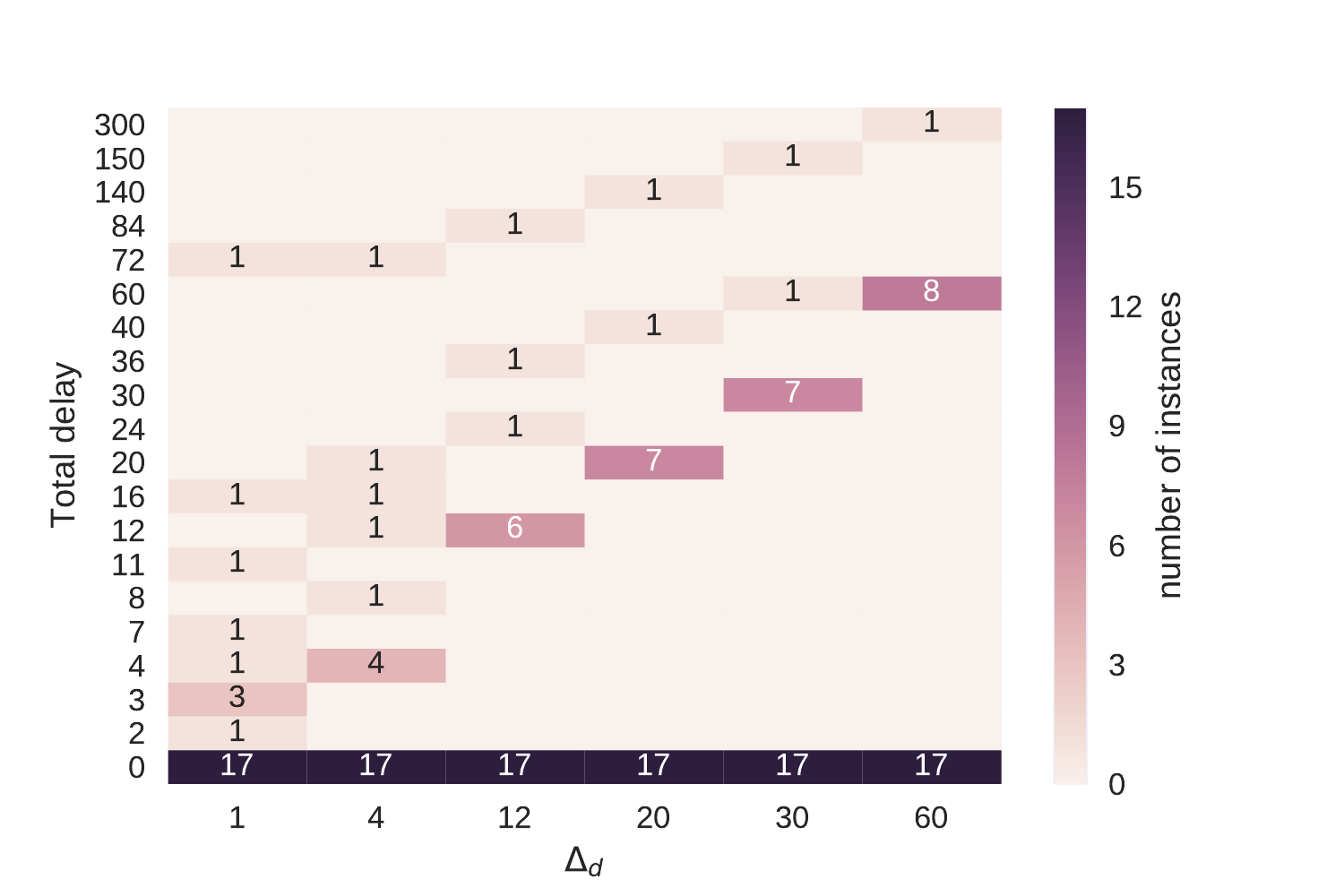}
    \caption{Total delay in dependence of the discretization parameter
    $\Delta_d$ for 26 different problem instances from $\mathcal{I}_{60}$ with
    up to in $12$ flights and $25$ conflicts. 
    The color code shows the number of instances with the same total delay.
    17 of these 26 instances had trivial solutions for all values of $\Delta_d$, i.e the total delay vanishes.
    }
\label{fig:icm1}
\end{figure}

Figure~\ref{fig:icm1} shows the dependence of the total delay time optimized
by ICM on the delay discretization $\Delta_d$ for various problem instances
extracted from the connected components of the conflict graph.  Results are for
maximum delay of 60 minutes.  As expected, the total delay decreases by
decreasing $\Delta_d$.  This is consistent with the idea that smaller
$\Delta_d$ allows a finer optimization of the delays of the flights.

\begin{figure}[h]
  \includegraphics[width=\columnwidth]{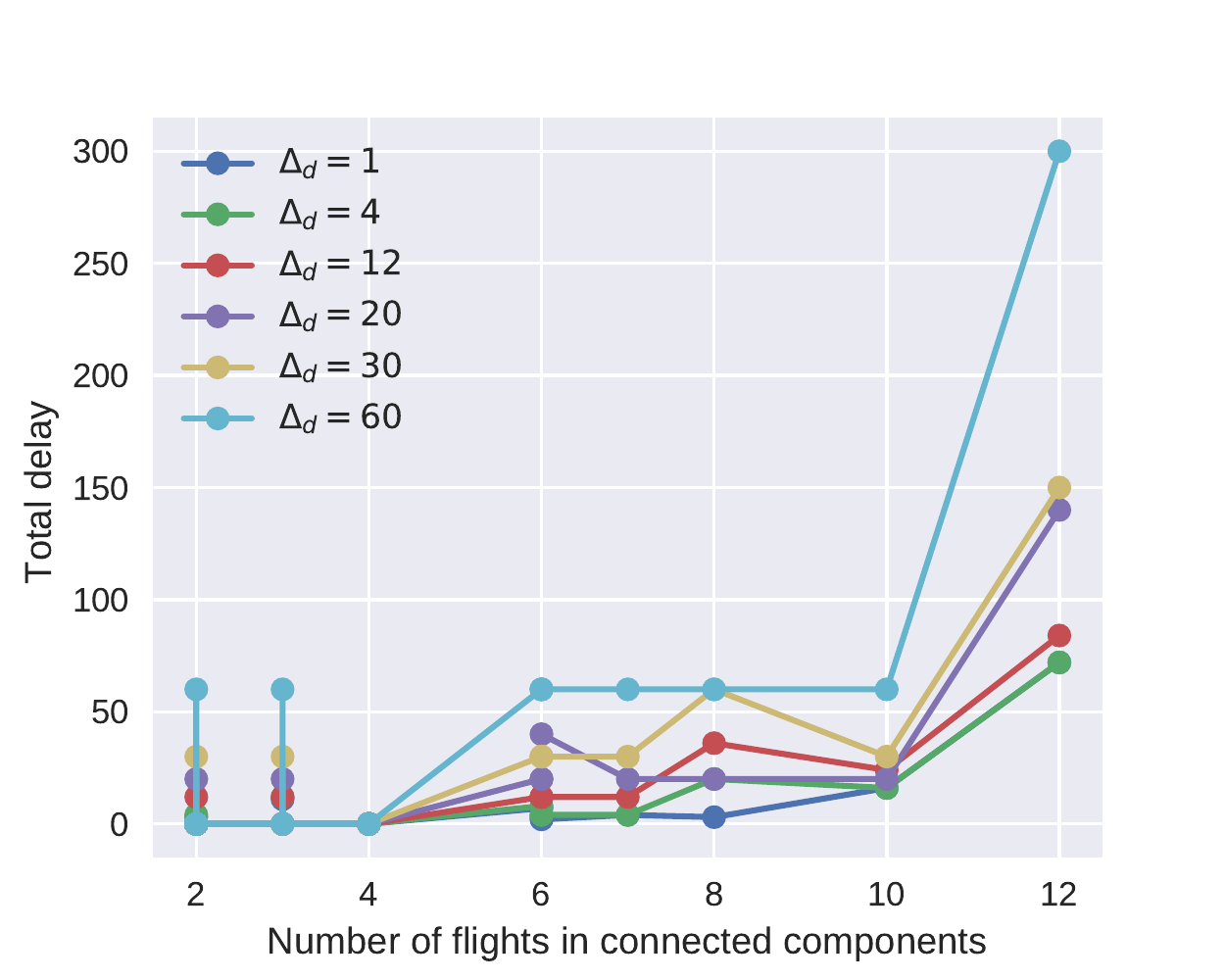}
  \caption{\label{fig:icm2}. Optimal total delay found by using the
  Isoenergetic Cluster Method (ICM) at fixed time step $\Delta_d$ as a function
  of numbers of flight within each connected component. ICM was unable to find
  solutions for connected component with more than $12$ flights.}
\end{figure}

In figure~\ref{fig:icm2} we show the optimal delay time found by ICM as a
function of the number of the flights in the connected components. Results are
for a maximum delay of 60 minutes. Unfortunately, ICM was unable to optimize
connected components with more than $12$ flights. This can be explained by
recalling that ICM works the best for almost-planar problem while its
performance quickly decreases for fully-connected problems. Indeed, as shown in
Section~\ref{sec:instances}, the underlying graph of connected components look
more and more like a fully-connected graph rather than a tree graph as the
number of flights inside the connected component increases.

%%%%%% MAPPING %%%%%%
\section{Mapping to QUBO}
\label{sec:mapping}
In this section, we describe how to map to QUBO from the conflict-resolution problem
limited to only departure delays; a more general mapping is given in the
appendix.

\subsection{Binary encoding}
Having suitably discretized the configuration space, we must then encode it
into binary-valued variables.  The value of $d_i$ is encoded in $N_d + 1$
variables $d_{i,0}, \ldots, d_{i,N_d} \in \BB$ using a one-hot encoding:
\begin{equation}
\label{eq:d-encoding}
d_{i, l} = \begin{cases}
1, & d_i = l,\\
0, & d_i \neq l;
\end{cases}
\qquad
d_i = \Delta_d \sum_{l = 0}^{N_d} l d_{i,l}.
\end{equation}
To enforce this encoding, we add the penalty function
\begin{equation}
\label{eq:dep-delay-encoding-penalty}
\function{encoding} = 
\weight{encoding} 
\sum_{i = 1}^{\Nf} 
{\left(
\sum_{l = 0}^{N_d} d_{i,l} - 1
\right)}^2,
\end{equation}
where $\weight{encoding}$ is a penalty weight sufficiently large to ensure that any cost minimizing state satisfies $\function{encoding} = 0$.
(Note that in practice, we could do away with the bit $d_{i,0}$ by removing it from~\eqref{eq:d-encoding} and substituting 
$\sum_{l=1}^{N_d - 1} \sum_{l'=l+1}^{N_d} d_{i,l} d_{i,l'}$
into $\function{encoding}$.)
In terms of these binary variables, the total delay contribution to the cost function is 
\begin{equation}
\label{departure_delay_model_qubo_departure}
\function{delay} = 
\Delta_d
\sum_{i=1}^{\Nf}
\sum_{l=0}^{N_d} l d_{i,l},
\end{equation}
Lastly, actualized conflicts are penalized by 
\begin{equation}
  \function{conflict}
=
\weight{conflict}
\sum_{k}
\sum_{\substack{\left.l, l' \middle| \Delta_d (l - l') \in D_k\right.\\
\left. i, j \in I_k \middle| i < j \right.}}
  d_{i,l} d_{j, l},
\end{equation}
where again $\weight{conflict}$ is a sufficiently large penalty weight. 
The overall cost function to be minimized is 
\begin{equation}
f
=
\function{encoding} + \function{delay} + \function{conflict}.
\end{equation}

%%%%%% SOFT CONSTRAINTS %%%%%%
\subsection{Softening the constraints}
In the QUBO formalism, there are no hard constraints; thus we use of penalty
functions in the previous section.  For sufficiently large penalty weights, the
optimum will satisfy the desired constraints.  However, precision is a limited
resource in quantum annealing; therefore, we would like to determine the
smallest sufficient penalty weights, at least at the level of the classical model~\cite{zhu2016best}.

\begin{figure}[htpb]
\centering
\includegraphics[width=0.45\textwidth]{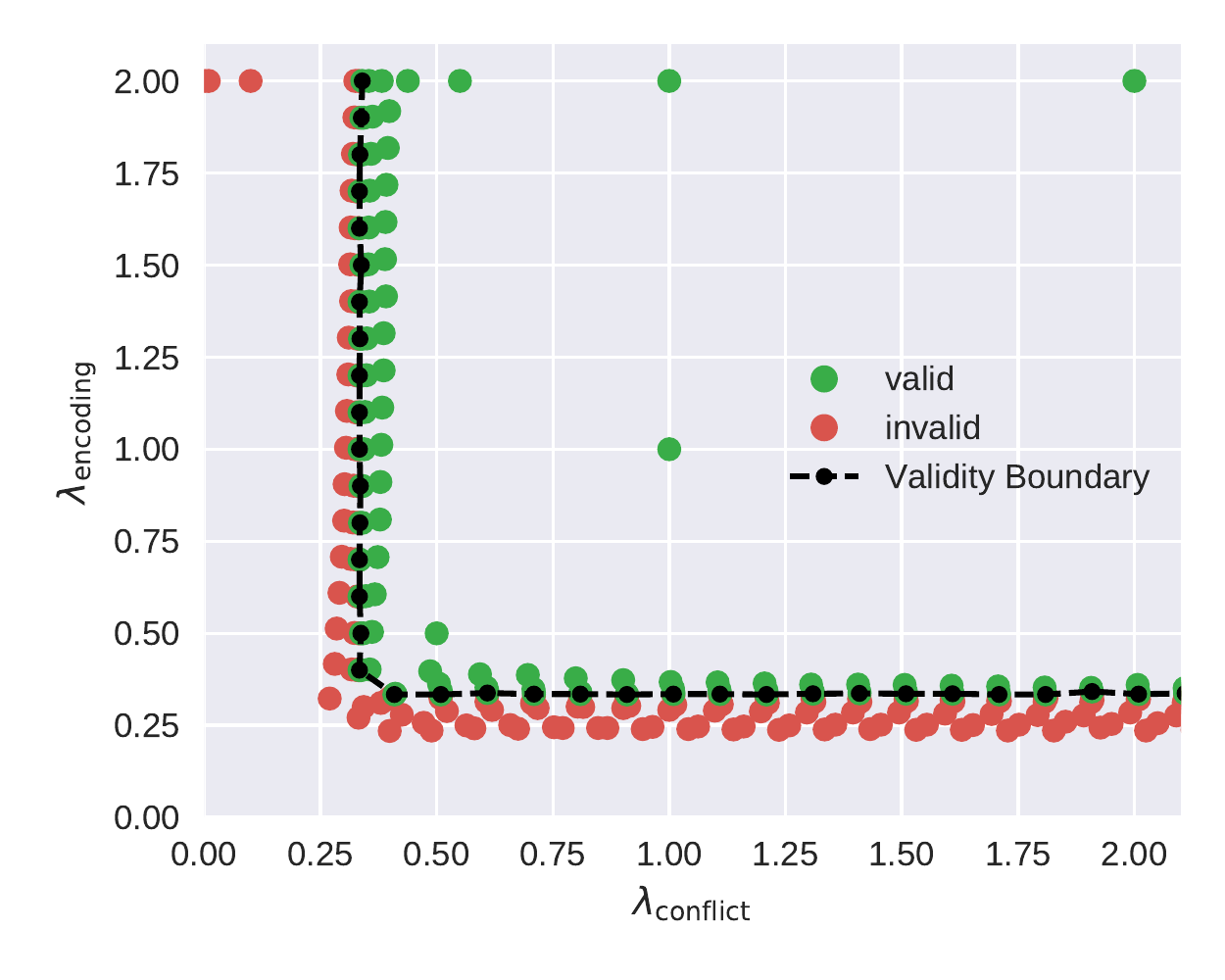}
\caption[Penalty weight phase diagram]{Validity of exact solution to a QUBO
extracted from a problem instance with $N_f=7$ flights and $N_c=9$ conflicts in
dependence on the choice of the penalty weights, $\lambda_\text{encoding}$ and
$\lambda_\text{conflict}$. Here, $\Delta_t=6$ and $d_\text{max}=18$.
In order to obtain the exact solutions, we used a Max-SAT solver~\cite{akmaxsat} after we mapped the QUBO instances to Max-SAT instances.
}
\label{fig:penalty_weights}
\end{figure}

For a given instance, we say that a pair of penalty weights
$(\weight{conflict}, \weight{encoding})$ is valid if the minimum of the total
cost function satisfies both the conflict and encoding constraints when using
those weights.  Figure~\ref{fig:penalty_weights} shows the phase space of these
penalty weights for a single instance with $7$ flights and $9$ conflicts.  The
box-like boundary between valid and invalid penalty weights suggests that the
validity of the two penalty weights is independent; this box-like boundary is
found for all of our instances with up to $7$ flights and $9$ conflicts.

%%%%%% D-WAVE RESULTS %%%%%%
\section{Quantum Annealing}
\label{sec:qa}
In this section we report on our efforts to solve problem instances from the
departure delay model from Section~\ref{sec:mapping} with a D-Wave 2X quantum
annealer.  We restricted ourselves to instances with
$d_\text{max}=D_\text{max}=18$ and $\Delta_d \in \{3, 6, 9\}$.

\subsection{Background}
Quantum annealing (QA) is a heuristic for minimizing pseudo-Boolean functions $f: {\{1, -1\}}^n \to \mathbb R$ using specially-designed quantum hardware.
In physical terms, the goal is to find a ground state (i.e.\ lowest-energy state) of the ``problem Hamiltonian''
\begin{equation}
\hat{H}_f = \sum_{i} h_i \hat{Z}_i + \sum_{i,j} J_{i, j} \hat{Z}_i \hat{Z}_j,
\end{equation}
where $\hat{Z}_i$ is the Pauli Z operator acting on the $i$th qubit and 
\begin{equation}
f(\mathbf s) = 
\sum_{i} h_i s_i + \sum_{i,j} J_{i, j} s_i s_j
\end{equation}
is the unique multilinear form of $f$.
This is done by starting in a uniform superpositon of the computational basis states, which is the ground state of the initial Hamiltonian
\begin{equation}
    H_0 = \sum_{i} \hat{X}_i,
\end{equation}
where $\hat{X}_i$ is the Pauli X operator acting on the $i$th qubit.
The adiabatic theorem implies that if we change the system's Hamiltonian from the initial one to the final one slowly enough, then at all times the system will remain in its ground state, including at the end, which yields the ground state of the final Hamiltonian that we want.
The essential principle is that excitations to higher-energy states are suppressed to a degree related by their difference in energy from the ground state.
In practice, system noise and other factors mean that this ideal is not achieved, but often a low-energy state is a obtained, i.e.\ practical QA only \emph{approximately} minimizes the function $f$.

One obstruction to applying a particular quantum annealer to a given function is that the pairs of qubits $\{i, j\}$ for a which a term $J_{i,j}\hat{Z}_i \hat{Z}_j$ can be included in the problem Hamiltonian $H_f$ are restricted.
These restrictions are captured in what we'll call the ``hardware graph'', whose vertices correspond to qubits and whose edges indicate for which pairs of qubits a term in the Hamiltonian can be included.
Similarly, the vertices of the ``problem graph'' correspond to the Boolean variables $\mathbf s$ and the edges to the quadratic terms in the multilinear expansion.
Because the hardware graph is inherently fixed and bounded-degree, the problem graph is usually not a subgraph thereof, meaning that we cannot directly assign each variable $s_i$ to a single qubit.
This is addressed by ``graph-minor embedding'', in which each variable is mapped to a set of qubits. 
This is done is such a way that when the vertices of the hardware graph corresponding to each vertex of the problem graph are contracted into one, the resulting graph is isomorphic to the problem graph.
Lastly, additional terms, called ``intra-logical couplings'' are added between the qubits that each variable is mapped to in order to ensure that they act as one, by penalizing states in which the state of those qubits are not the same.

\subsection{Embedding}
In order to make a QUBO amenable for a D-Wave 2X quantum annealer, it has to
obey certain hardware constraints.  For instance the connections between the
binary variables are restricted to the so called Chimera
graph~\cite{Rieffel2015}.  However, it is possible to map every QUBO to another
QUBO which obeys the constraints of the Chimera architecture while increasing
the number of binary variables used by a so called minor-embedding
technique~\cite{choi}

\begin{table}[h]
    \begin{tabular}{lrrr}
    \toprule
    $\Delta_d$ &    3 &    6 &    9 \\
    \midrule
    Number of flights $N_f$   &   13 &   19 &   50 \\
    Number of conflicts $N_c$ &   27 &   47 &  104 \\
    Number of logical qubits  &   91 &   76 &  150 \\
    Average number of physical qubits &  631 &  395 &  543 \\
    \bottomrule
    \end{tabular}
    \caption{Parameters of the largest embeddable instances for the D-Wave 2X}
\label{tab:embedding}
\end{table}

\begin{table}[h]
    \begin{tabular}{lrrr}
    \toprule
    $\Delta_d$ &    3 &    6 &    9 \\
    \midrule
    Number of flights $N_f$   &   19 &   50 &   64 \\
    Number of conflicts $N_c$ &   47 &   104 &  261 \\
    Number of logical qubits  &  133 &   200 &  192 \\
    Average number of physical qubits &  1235 &  1080 &  1121 \\
    \bottomrule
    \end{tabular}
    \caption{Parameters of the largest embeddable instances for the D-Wave 2000Q}
\label{tab:embedding_D2000Q}
\end{table}

\begin{figure}[htpb]
\centering
\includegraphics[width=0.45\textwidth]{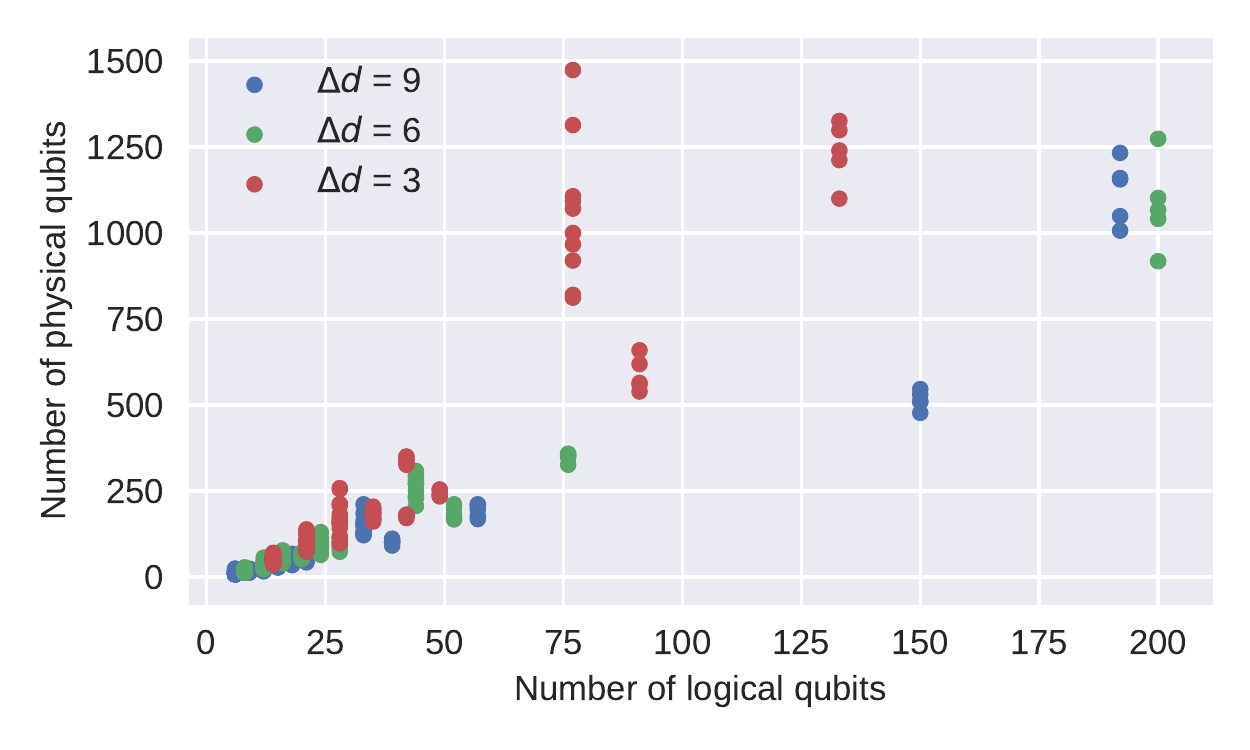}
\caption{Number of physical qubits versus the number of logical qubits for embeddings into a D-Wave 2000Q of the QUBO instances corresponding to $\mathcal I_{18}$.}
\label{fig:number_of_physical_qubits}
\end{figure}

Of course the QUBO graph structure of the instances is not suitable for direct calculation on the
D-Wave machine. Therefore we used D-Wave's heuristic embedding
algorithm~\cite{DWaveEmbedding} to embed instances with up to $N_f=50$
and $N_c=104$ on the D-Wave 2X as well as up to $N_f=64$ and $N_c=261$ on the
D-Wave 2000Q depending on discretization (cf.\ Tables~\ref{tab:embedding} and
\ref{tab:embedding_D2000Q}). We generated up to $5$ different embeddings
for each QUBO instance, and selected the one that used the smallest number of
physical qubits.  In figure~\ref{fig:number_of_physical_qubits} one can see the
dependence of the number of physical qubits on the number of logical qubits.

%%%%%% SUCCESS PROBABILITY %%%%%%
\subsection{Success Probability}
In order to investigate the performance of the D-Wave machines 2X and 2000Q, we
compared the annealing results to the ones of an exact solver.  We used an
exact Max-SAT solver~\cite{akmaxsat} after we mapped the QUBOs to Max-SAT.  For
each QUBO instance, we ran the annealing process in between $10^4$ and $10^6$
times.  The success probability $p$ is then given by the ratio of the number of
annealing solutions which are equal to the exact solution and the number of
total annealing runs.  As a measure of the runtime of the machine, we used the
time-to-solution with probability 99\%.
\begin{equation*}
    T_{99} = \frac{\ln(1 - 0.99)}{\ln(1 - p)} T_\text{Anneal} \, ,
\end{equation*}
where $T_\text{Anneal}$ is the annealing time which was set to $20 \mu s$.

\begin{figure}[htpb]
    \centering
    \includegraphics[width=0.45\textwidth]{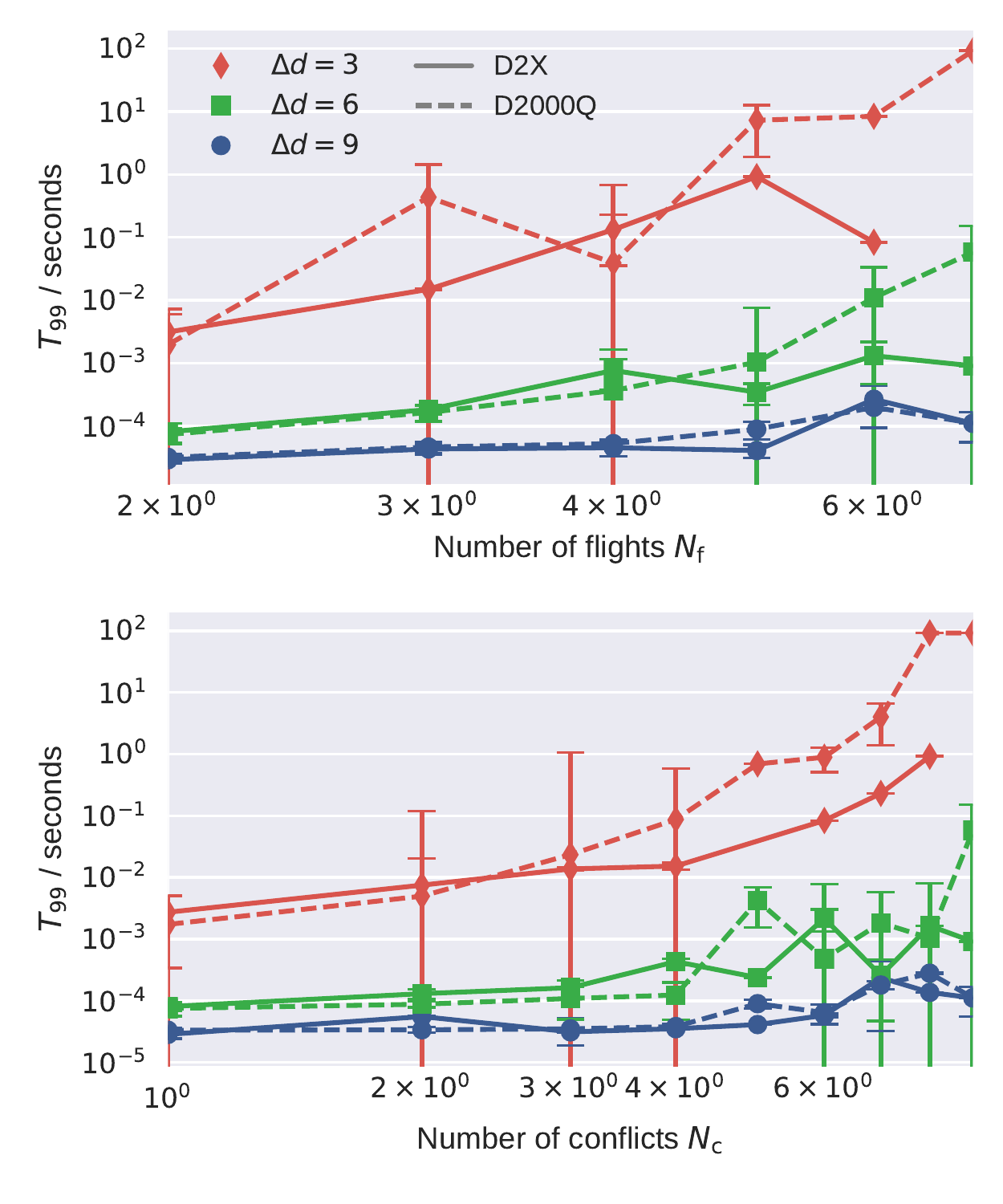}
    \caption{Median of the time to solution with 99 \% with probability
    $T_\text{99}$ for QUBO instances in dependence of the number of flights
    $N_f$ and the number of conflicts $N_c$.  The error bars indicate the $35
    \%$ and $65 \%$ percentiles.  We used $10000$ annealing runs for each
    instance, penalty weights $\lambda_\text{conflict} =
    \lambda_\text{encoding} = 1$ and $5$ different embeddings.  The
    ferromagnetic coupling between all physical qubits of the same logical
    qubit was set to $J_F=-1$ in absolute units for all the instances.  The
    solid lines are results from the D-Wave 2X whereas the dashed lines are
    results from the D-Wave 2000Q. For these results, we did not use gauges and used energy minimization to deal with broken qubit chains.
    }
\label{fig:time_to_solution}
\end{figure}

In figure~\ref{fig:time_to_solution} the dependence of the time to solution
$T_{99}$ on the number of flights and the number of conflicts is shown.  One can
see, that the success probability decreases for larger problem instances as
well as for finer discretizations.  We conjecture, that this is mainly due to
the limited precision in the specification of a QUBO on the D-Wave machines.
In order to investigate the influence of limited precision, we need a measure
for the precision needed to represent a given QUBO instance.
If the embedded QUBO instance is given by $H = \sum_{ij} Q_{ij} x_i x_j$ with $x_i \in \{0, 1\}$ the corresponding Ising model
\begin{equation*}
    H = \sum_i h_i s_i + \sum_{ij} J_{ij} s_i s_j, \qquad s_i \in \{-1, 1\} \, ,
\end{equation*}
can be obtained by the transformation $s_i = 2 x_i - 1$.
A measure for the precision needed is then given by the \emph{maximum coefficient ratio}
\begin{equation*}
    C_\text{max} = \max\left[\frac{\max_i | h_i |}{\min_i | h_i |} ,\frac{\max_{ij} | J_{ij} |}{\min_{ij} | J_{ij} |} \right] \, .
\end{equation*}
The larger this number is, the finer precision is needed for correctly representing the QUBO on a D-Wave machine.

\begin{figure}[htpb]
    \centering
    \includegraphics[width=0.45\textwidth]{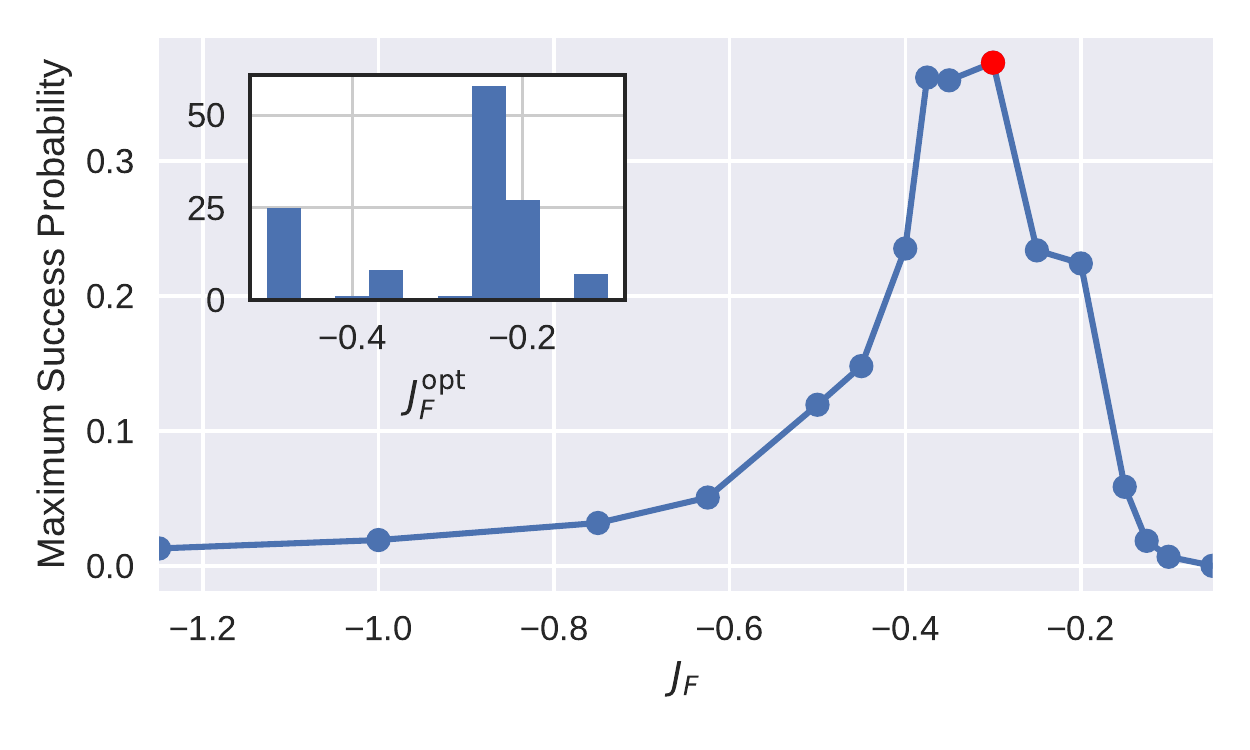}
    \caption{Maximum success probability on the D-Wave 2000Q for a QUBO instance with $N_f=5$, $N_c=5$ and $\Delta_d=6$ in dependence of $J_F$,
             where $J_F$ is given in units of the largest coefficient of the embedded Ising model.
             We used 5 different embeddings, $100000$ annealing runs and penalty weights $\lambda = \lambda_\text{conflict} = \lambda_\text{unique} =1$ for each of the data points.
             The red data point indicates the optimal value $J_F^\text{opt}$.
             The inset shows the distribution of $J_F^\text{opt}$ for all solvable instances.
    }
\label{fig:success_probability_vs_jferro}
\end{figure}

\begin{figure}[htpb]
    \centering
    \includegraphics[width=0.45\textwidth]{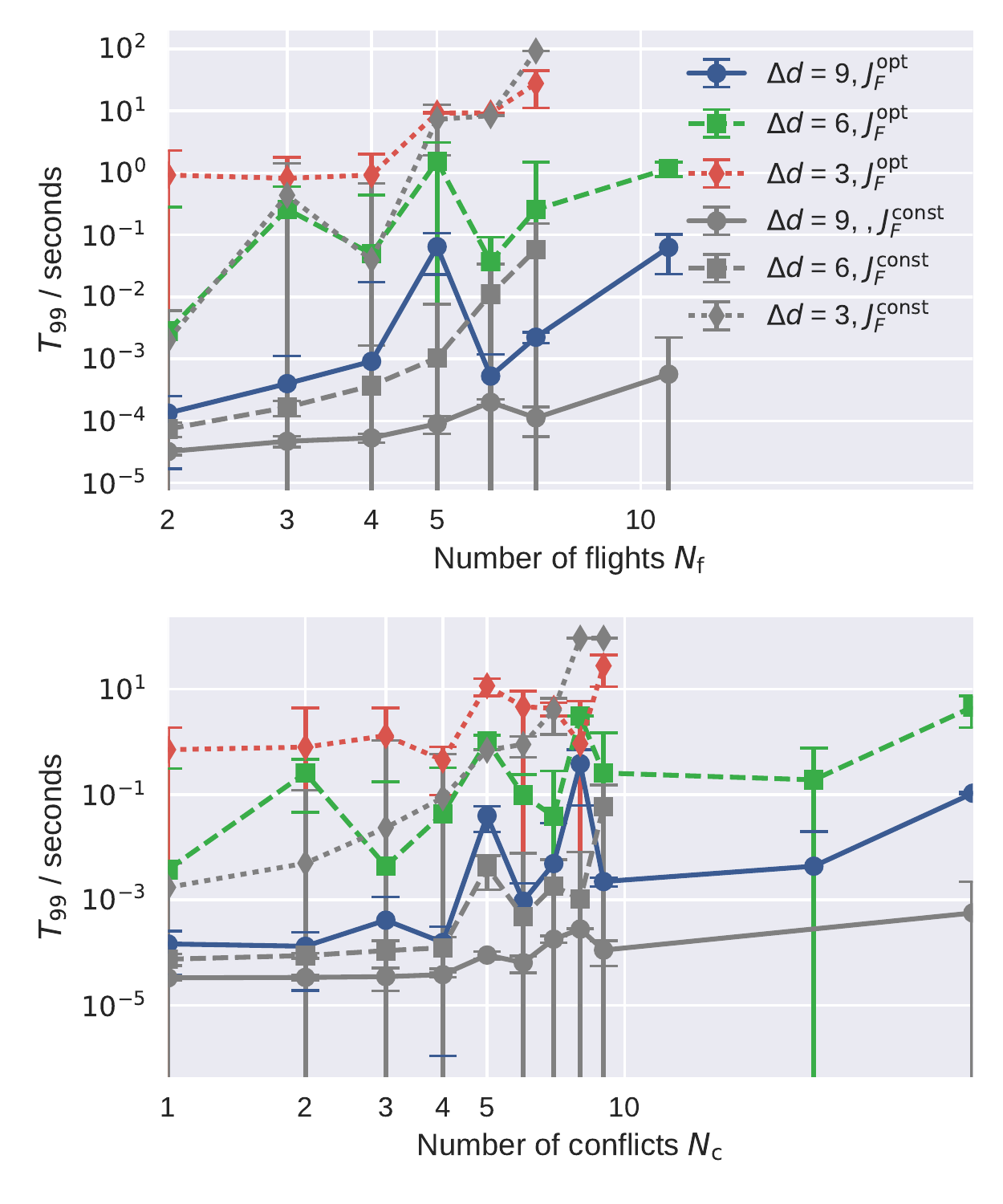}
    \caption{Median of the time to solution with 99 \% with probability $T_\text{99}$ for fixed and optimal $J_F$.
         The error bars indicate the $35 \%$ and $65 \%$ percentiles.  We used
         up to $1000000$ annealing runs on the D-Wave 2000Q for each instance,
         penalty weights $\lambda_\text{conflict} = \lambda_\text{encoding} =
         1$ and up to $5$ different embeddings.  The colored lines indicate
         results obtained with an optimal $J_F=J_F^\text{opt}$ which is
         instance dependent.  The grey lines indicate results obtained with a
         fixed $J_F=J_f^\text{const}=-1$ in absolute units.
    }
\label{fig:time_to_solution_optimal_jferro}
\end{figure}

The success probability also depends on the choice of the ferromagnetic
intra-logical qubit coupling $J_F$.  
Figure~\ref{fig:success_probability_vs_jferro} shows the dependence of the success
probability on this coupling for one particular problem instance.  However, the
general behavior of this curve is instance independent.  For very small $J_F$,
the qubit strings which represent logical qubits might be broken by coupling to
outer logical qubits and the success probability is suppressed.  On the other
hand, if $J_F$ is very large, the precision needed $C_\text{max}$ will
eventually surpass the machine precision and the success probability will
decrease.  In between these two extrema, there will be a sweet spot with an
optimal $J_F^\text{opt}$ which yield maximum success probability.  We
determined the optimal coupling $J_F^\text{opt}$ for the problem instances by
sweeping over values in between $J_F=-1.25$ to $J_F=-0.125$ in units of the
largest coefficient of the embedded Ising model.  The inset in figure
\ref{fig:success_probability_vs_jferro} shows the distribution of the
$J_F^\text{opt}$.  Using the optimal couplings, the performance is increased
with respect to a constant value $J_F^\text{const} = -1$ in absolute units as
one can see in figure \ref{fig:time_to_solution_optimal_jferro}.

\begin{figure}[htpb]
    \centering
    \includegraphics[width=0.45\textwidth]{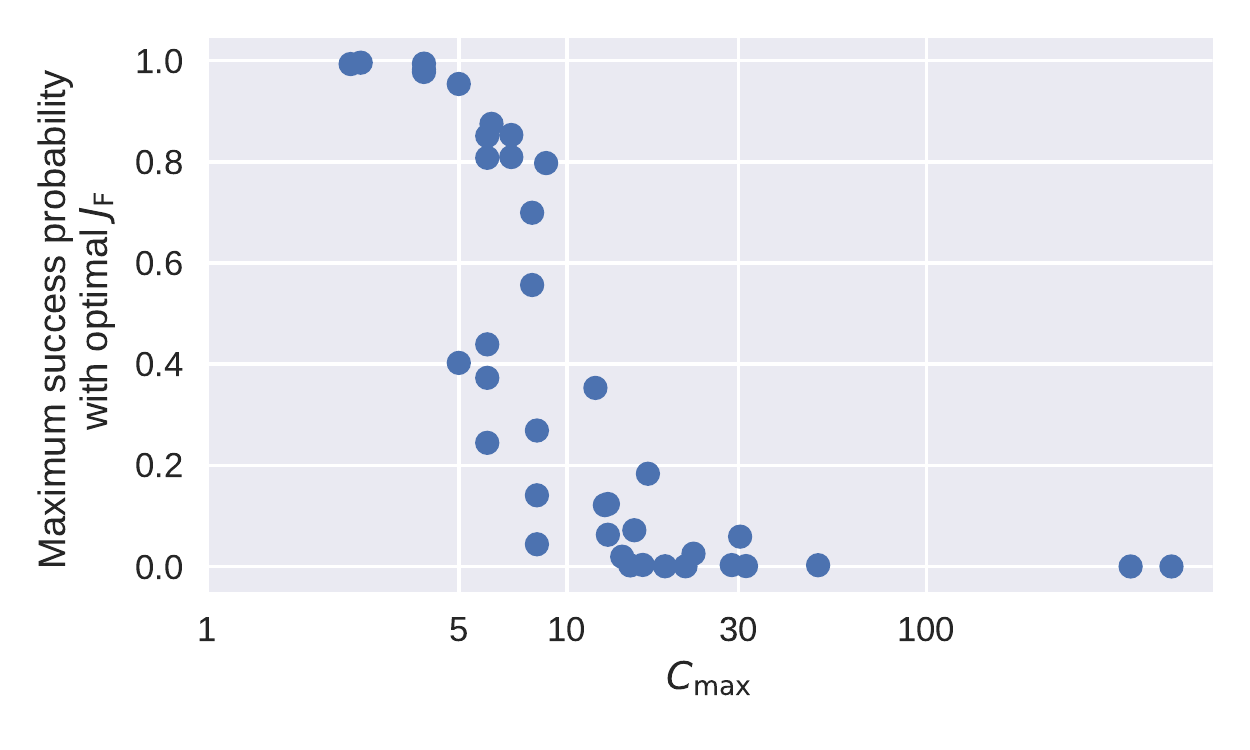}
    \caption{Maximum success probability on the D-Wave 2000Q for optimal $J_F$ for all embeddable instances in $\mathcal{I}_{18}$ against the coefficient ratio $C_\text{max}$.
    }
\label{fig:success_probability_vs_cmax}
\end{figure}

Using the optimal coupling $J_F^\text{opt}$ we can study the influence of the
limited machine precision on the success probability.  Figure
\ref{fig:success_probability_vs_cmax} shows the maximum success probability
with optimal $J_F$ for all embeddable instances in $\mathcal{I}_{18}$.  The
influence of the limited machine precision can be seen in the decrease of the
success probability with increasing precision $C_\text{max}$.  The success
probability vanishes around $C_\text{max}\approx 30$ which corresponds to the
machine precision of the D-Wave 2000Q of around $\sim 1/30$.  Since
$C_\text{max}$ in general increases with the problem size as well as with finer
discretizations, this explains the long time-to-solutions for large instances
and fine discretizations in figures \ref{fig:time_to_solution} and
\ref{fig:time_to_solution_optimal_jferro}.

%%%%%% CONCLUSIONS %%%%%%
\section{Conclusions}
Quantum annealing is a relatively new heuristic that offers hope of solving classical optimization problems better in several ways compared to classical alternatives:
higher-quality solutions, faster time to solution, many approximate solutions, and qualitatively different solutions.
Whether this hope will likely have to be answered empirically, by running problems on actual hardware.
This work is a first step towards that end, but not the last.
In particular, the restriction to time delays only is obviously unrealistic.
Nevertheless, this simplification both serves a proof of principle and enables runs on extant quantum annealers at a reasonable problem scale.
To be practically relevant, this line of work must be extended to include in the QUBO all practically relevant aspects of the problem.
The full problem can then be solved using other promising QUBO solvers such as ICM, which may improve on the state of the art even if quantum annealing does not.

In this paper, we propose a novel QUBO mapping for a simplified version of the
Air Traffic Management (ATM) conflict-resolution problem for wind-optimal
trajectories involving minimum trajectory modifications. Although these efforts
are driven by making the problem amenable to quantum annealers, the techniques
used may be beneficial also for the classical solution of the problem.  In our
study, we considered the actual wind-optimal trajectories for transatlantic
flights (NAT) on July 29, 2012. Given the large number of flights, the
wind-optimal trajectories cannot be directly mapped in a QUBO model.  To
overcome this limitation, our modified version of the conflict-resolution
problem assumes that the flight maneuvers applied to avoid conflicts modify the
wind-optimal trajectories only locally, resulting in assigning ``delays'' to
the flights. Therefore, wind-optimal trajectories can be ``hard encoded'' in
our QUBO formulation of the conflict-resolution problem with the flights delays
being the only variables to optimize. Nevertheless, as explained in Appendix 2,
our method is general enough to potentially include the effect of other
maneuvers as well.

As part of our study, we also introduce a novel ``pre-processing'' algorithm to
eliminate potential conflicts that, given a maximum delay, can never occur and clustering adjacent conflicts.
This novel approach does not only reduce the number of potential conflicts, but is also gives an important
indication of the underlying topology the conflict graph. Indeed, we have
discovered that most of the flights have very few conflicts while there are few
flights that have conflicts in a non trivial way.
The latter sets of flights
represent the hardest part of the conflict-resolution problem to optimize. We
want to emphasize that the proposed pre-processing algorithm is general and can
be successfully applied to the existing conflict-resolution methods to improve
both the speed and quality of solutions.

We also present several different QUBO mapping including local and global trajectory
deviations as well as including and excluding maneuvers.  Due to the hardware
limitations of the D-Wave machine we focus on a model excluding maneuvers in
order to keep the number of variables small.  Using D-Wave's embedding
algorithm, several smaller problem instances were embeddable onto the D-Wave 2X
as well as onto the D-Wave 2000Q.  However, the limiting factor for the success
probability is not the sizes of the chips but its limited precision.  Therefore
the success probability is suppressed for finer discretizations and larger
problem sizes.

Finally, we have analyzed the performance of both classical and quantum
heuristics in solving the QUBO model where only delays at the departure are
allowed. Results show that it is already hard to find conflict-free solution
for a flight set that involve more than 12 flights.

This work represents the foundation for future work, including:
\begin{itemize}
    \item Embed and solve QUBO instances for models that benefits from variable
          simplification (see section \ref{sec:mapping}), also including maneuvers on
          a quantum annealer.
    \item Improve performance of quantum annealing by alternative embedding
          strategies and advanced annealing schedules available on newer D-Wave
          devices.
    \item Use best-available classical solvers that exploit the conflict graph
          description for classical solutions to the problem. The studied problem is
          related to several works in multi-agent path
          planning~\cite{akella2002coordinating,wang2015collision} and no-wait
          job-shop-scheduling problems~\cite{sahni1979complexity} and the quantum
          annealing results could be benchmarked with the domain-specific solvers of
          these related problems, after careful mapping.
\end{itemize}

%%%%%% ACKNOWLEDGMENTS %%%%%%
\section{Acknowledgements}

This work has been funded by the NASA Aviation Systems Division.
D.V. was supported by NASA Academic Mission Services, contract number NNA16BD14C.
The authors would like to acknowledge additional support from the NASA Advanced Exploration Systems program
and the NASA Ames Research Center. The views and conclusions contained herein are
those of the authors and should not be interpreted as necessarily representing the official
policies or endorsements, either expressed or implied, of the U.S. Government. The U.S.
Government is authorized to reproduce and distribute reprints for Governmental purpose
notwithstanding any copyright annotation thereon.

\appendix

%%%%%% APPENDIX: MAPPING %%%%%%
\section{General QUBO mapping}
In this section we describe a mapping to QUBO of a more general version of the deconflicting problem than that covered in the main text.

\subsection{Alternative encodings}
In the mappings describe both in the main text and the appendix, we use a one-hot encoding to encode a variable.
This is best for the specific mappings we described, but in variants an alternative may be better.
Say we have a variable $x$ that we want to allow to have variables from finite set $W = \{w_1, w_2, \ldots, w_m\}$.
The one-hot encoding has $m$ bits ${\left(x_{i}\right)}_{i=1}^m$ such that $x = \sum_{i=1}^m w_i x_i$ and $\sum_{i=1}^m x_i = 1$.
While we focus on the case in which $W = \{0, 1, \ldots, m -1\}$, our methods are not dependent on that being case, and in particular can address non-uniform sets of values, say if via clever preprocessing it can be determined that such a set would be sufficient.
An alternative encoding would remove the requirement that exactly one of the bits is one.
The variable $x$ would still be encoded as $x = \sum_{i=1}^m w_i x_i$, but without the one-hot constraint can take on values in $\left\{\sum_i b_i w_i \middle| b_i \in \{0, 1\}\right\}$.
In particular, this encompasses the unary encoding in which $w_i = 1$ for all $i$ and thus $x \in [0, m]$, as well as the binary encoding $w_i = 2^{i-1}$ for which $x \in [0, 2^m - 1]$.
The latter has the advantage of requiring much fewer qubits, but at the cost of similarly increased precision.
The former requires the same number of qubits as the one-hot encoding we use, and even has the benefit of minimal precision, but does not allow for quadratic constraints that penalize certain pairs of values of variables, e.g. $d_i - d_j \neq B_k$, without the use of ancillary bits.
In models in which the bits $x_{i}$ only appear in the sum $\sum_i x_i$, it is actually preferable to use the unary encoding to improve the precision requirements.
We stick to the one-hot encoding for simplicity, but in practice the unary encoding should be used when possible.

To make the expressions more concise, we define the generalized encoding penalty function
\begin{equation}
\function{encoding}
\left(
{\left\{X_i\right\}}_i
\right)
=
\weight{encoding}
\sum_i
{\left(
  \sum_{x \in X_i} x - 1
\right)}^2
\end{equation}
that enforces the constraint that exactly one bit $x$ is one for each set of bits $X_i$.

\subsection{Global trajectory modifications}
Consider the case in which each trajectory can be modified by a departure delay and some parameterized spatial transformation, i.e.\ for each flight $i$ there is a variable $d_i$ and some parameter $\boldsymbol \theta_i$.
For example, Rodionova et al.~\cite{rodionova16} consider a single angle $\theta_i$ that determines a sinusoidal transformation of the trajectory.
For the QUBO mapping, we require that these variables be allowed to take on values from some finite set, so that are QUBO variables are $\{d_{i, \alpha}\}$ and $\{\boldsymbol \theta_{i,\phi}\}$, where $d_{i, \alpha} = 1$ ($d_{i, \alpha}$) indicates that $d_i = \alpha$ ($d_i \neq \alpha$) and similarly for $\boldsymbol \theta_{i, \phi}$.
For every pair of flights $i < j$, we can efficiently (in time and space polynomial in the size of the input) compute whether the corresponding trajectories conflict when modified according to $d_i$, $d_j$, $\boldsymbol \theta_i$ and $\boldsymbol \theta_j$.
Let $B_{i, j}$ be the set of values of $(d_i, \boldsymbol \theta_i, d_j, \boldsymbol \theta_j)$ such that the the modified trajectories conflict.
Lastly, let $d_{(i, \alpha), (j, \beta)}=1$ indicate that $d_i = \alpha$ and $ d_j = \beta$, and similarly for $\boldsymbol \theta_{(i, \phi), (j, \psi)}$.
The overall cost function is
\begin{multline}
\function{global}
\left(
{\left(d_{i,\alpha}\right)}_{i,\alpha}
{\left(d_{(i,\alpha),(j,\beta)}\right)}_{i,j,\alpha,\beta}
{\left(\boldsymbol \theta_{(i,\phi),(j,\psi)}\right)}_{i,j,\phi,\psi}
\right)
=\\
\function{encoding} +
\function{consistency} +
\function{delay} +
\function{conflict},
\end{multline}
where
\begin{equation}
\function{encoding} \left(
{\left\{
{\left\{
d_{i, \alpha}
\right\}}_{\alpha}
\cup
{\left\{
\boldsymbol \theta_{i, \phi}
\right\}}_{\phi}
\right\}}_i
\right)
\end{equation}

ensures that the values of $d_i$ and $\boldsymbol \theta_i$ are uniquely encoded;
\begin{multline}
\function{consistency}
=\\
\begin{split}
\weight{consistency}
\bigg[
&\sum_{i < j,\alpha, \beta}
s\left(d_{i,\alpha}, d_{j, \beta}, d_{(i, \alpha), (j, \beta)}\right)
\\
&+
\sum_{i < j,\phi, \psi}
s\left(\boldsymbol \theta_{i,\phi}, \boldsymbol \theta_{j, \psi}, \boldsymbol \theta_{(i, \phi), (j, \psi)}\right)
\bigg]
\end{split}
\end{multline}
ensures consistency between the values of $d_{i,\alpha}$, $d_{j, \beta}$, and $d_{(i, \alpha), (j, \beta)}$;
\begin{equation}
s(x, y, z) = 3z + xy - 2xz - 2 yz
\end{equation}
is a non-negative penalty function that is zero if and only if $z = xy$;
\begin{equation}
\function{delay}
= \sum_{i, \alpha} \alpha d_{i, \alpha}
\end{equation}
is the cost function to be minimized; and
\begin{equation}
  \function{conflict}
=
\weight{conflict}
\sum_{i < j}
\sum_{(\alpha, \phi, \beta, \psi) \in B_{i, j}}
d_{(i, \alpha), (j, \beta)} \boldsymbol \theta_{(i, \phi), (j, \psi)}
\end{equation}
penalize conflicts.

\subsection{Local trajectory modifications}

Alternatively, we can consider modifications to the trajectory only near conflicts.
We describe a few special models and their mapping to QUBO,
though many more such ways of doing so, and we leave a full accounting for future work.

\subsubsection{Exclusive avoidance}
Suppose for every conflict $k$ and associated pair of flights $i < j$, there is a way for either flight to go around the trajectory of the other, introducing some delay $d_{i,k}$ to flight $i$ or $d_{j, k}$ to flight $j$ depending on which trajectory is changed.
Let $a_k = a_{i, k} = 1$ ($a_{i, k} = 0$) indicate that flight $i$'s trajectory is changed (unchanged), and for convenience let $a_{j, k} = 1 - a_{i, k}$, though only one (qu)bit will be used per conflict. % chktex 36
Adding in the departure delay, we have the total cost function
\begin{equation}
\function{exclusive}
\left(
{\left(
  d_{i, \alpha}
\right)}_{i, \alpha},
{\left(a_k\right)}_k
\right)
=
\function{delay} +
\function{encoding},
\end{equation}
where
\begin{equation}
\label{eq:delay-exclusive}
\function{delay} =
\sum_{i}
\left[
\sum_{\alpha} \alpha d_{i, \alpha}
+
\sum_{k \in K_i} d_{i, k} a_{i, k}
\right]
\end{equation}
and
$\function{encoding}$ is as in~\eqref{eq:dep-delay-encoding-penalty}.
This assumes that the trajectory modifications don't introduce potential conflicts with other flights; this assumption can be partially relaxed by adding penalty terms of the form $a_{i,k} a_{j,k'}$ or $d_{i,\alpha} a_{j, k}$ as appropriate.

\subsubsection{Flexible avoidance}
In the exclusive avoidance model, it is \emph{required} that one or the other flight is delayed at each conflict.
We can relax this by accounting for the fact that if the flights arriving at a potential conflict are already relatively delayed, the conflict could be passively avoided (i.e.\ with no active maneuver).
Let $D_{k, \gamma} = 1$ ($D_{k, \gamma} = 0$) indicate that $D_k = \gamma$ ($D_k \neq \gamma$), where $D_k$ is the difference in the accumulated delays at conflict $k$ as defined in~\eqref{eq:accum-delay-diff}.

The total cost function is
\begin{multline}
\function{flexible}
\left(
{\left(d_{i, \alpha}\right)}_{i, \alpha},
{\left(a_{i, k}\right)}_{k, i \in I_k},
{\left(D_{k, \gamma}\right)}_{k, \gamma}
\right)
=\\
\function{encoding} +
\function{delay} +
\function{consistency} +
\function{conflict},
\end{multline}
where the first term is
\begin{equation}
\function{encoding}
\left(
{\left\{
{\left\{
d_{i, \alpha}
\right\}}_{\alpha}
\right\}}_i
\cup
{\left\{
{\left\{
D_{k, \gamma}
\right\}}_{\gamma}
\right\}}_{k}
\right);
\end{equation}

the consistency term is
\begin{equation}
\function{consistency}
=
\weight{consistency}
\sum_k
{\left(
D_{i, k} - D_{j, k}
-
\sum_{\gamma} \gamma D_{k, \gamma}
\right)}^2
\end{equation}
using the notational variables
\begin{equation}
D_{i, k} = \sum_{\alpha} \alpha d_{i, \alpha} +
\sum_{k' \in K_{i, k}}
d_{i, k'} a_{i, k'};
\end{equation}
$\function{delay}$ is as in~\eqref{eq:delay-exclusive} but where $a_{i,k}$ and $a_{j, k}$ are separate bits;
and
\begin{equation}
\function{conflict}
=
\weight{conflict}
\sum_k \sum_{\gamma \in B_k}
\left[
D_{k, \gamma}
\left(1 - a_{i, k} - a_{j, k}\right)
+ 2 a_{i, k} a_{j, k}
\right]
\end{equation}

If we want to allow both flights to be delayed at conflict $a_{i,k} = a_{j, k} = 1$, we must introduce an ancillary bit $a_k$ that indicates whether at least one flight is delayed at conflict $k$, adding
\begin{equation}
  \weight{consistency}
  \sum_{k}
  \left[
    \left(a_{i, k} + a_{j, k} \right) \left(1 - 2 a_k\right)
    + a_{i, k} a_{j, k}
  \right]
\end{equation}
to $\function{consistency}$, and
replacing $\function{conflict}$ with
\begin{equation}
\sum_k \sum_{\gamma \in B_k} D_{k, \gamma} (1 - a_k).
\end{equation}

\subsubsection{Interstitial delays}

In the interstitial-delay model, the local modifications are not made \emph{at} conflicts but \emph{between} them, and conflicts are only avoided via accumulated delays.
That is, the delay $d_{i, k}$ introduced to flight $i$ before reaching conflict $k$ but after leaving the previous conflict $\kappa_{i,k} = \max_{k' \in K_{i, k}}k'$.
Unlike in the flexible avoidance model, $d_{i,k}$ is now a variable rather than a parameter, and we encode it using bits $d_{i, k, \delta}$.
\begin{multline}
\function{interstitial}
\left(
{\left(
d_{i, \alpha}
\right)}_{i, \alpha},
{\left(
D_{i, k, \gamma}
\right)}_{i, k \in K_i, \gamma}
\right)
=\\
\function{encoding}
+
\function{consistency}
+
\function{conflict}
+
\function{delay},
\end{multline}
where
\begin{equation}
\function{encoding}
\left(
{\left\{
{\left\{
d_{i, \alpha}
\right\}}_{\alpha}
\right\}}_i
\cup
\bigcup_i
{\left\{
{\left\{
D_{i, k, \gamma}
\right\}}_{\gamma}
\right\}}_{k \in K_i}
\right),
\end{equation}

\begin{equation}
\function{consistency}
=
\sum_i
\sum_{k \in K_i}
\sum_{(\gamma, \gamma') \in B_{i, k}}
    D_{i, k, \gamma} D_{i, \kappa_{i,k}, \gamma'},
\end{equation}

\begin{equation}
\function{conflict}
=
\weight{conflict}
\sum_{k=1}^{\Nc}
\sum_{(\gamma, \gamma') \in B_k} D_{i, k, \gamma} D_{j, k, \gamma'},
\end{equation}
and
\begin{equation}
\function{delay}
\sum_i \sum_{\gamma} D_{i, \max K_i, \gamma}.
\end{equation}

\bibliography{atm}

\begin{IEEEbiography}[{\includegraphics[width=1in,height=1.25in,clip,keepaspectratio]{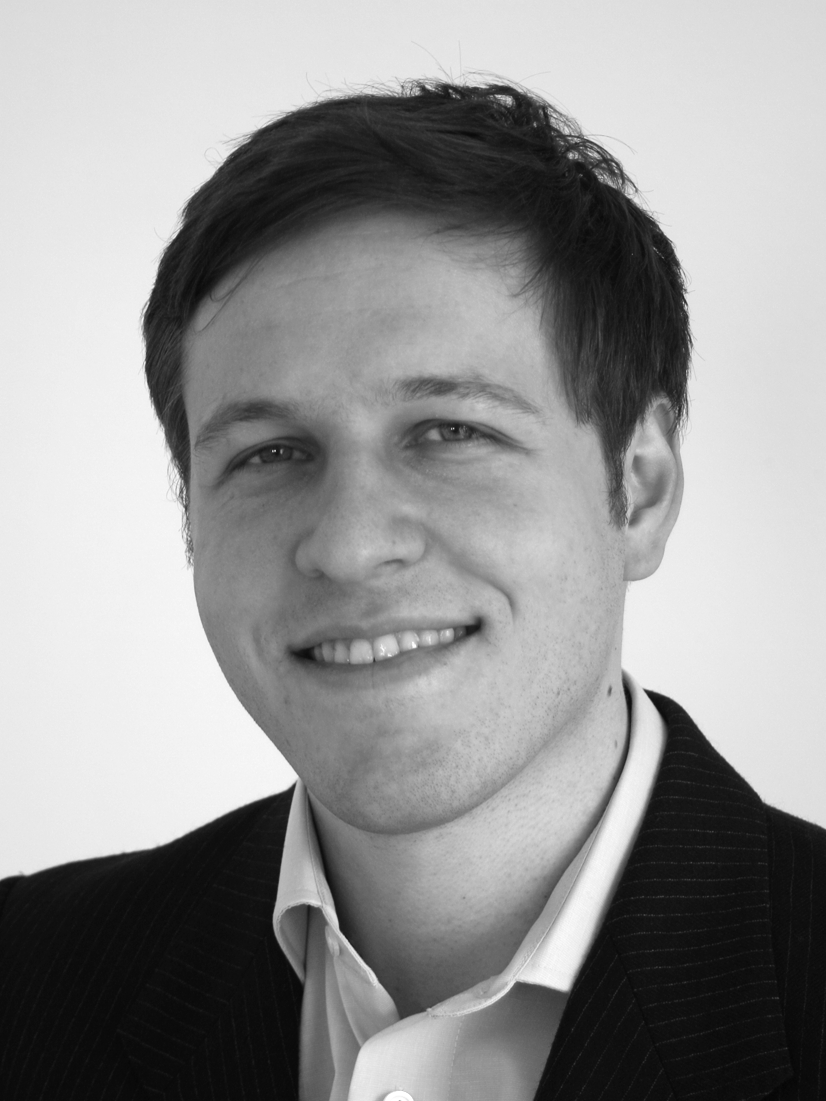}}]{Tobias Stollenwerk}
received his German diploma in physics in 2007 from the
University of Bonn. Afterwards he worked for two years as a research assistant
in computational photonics in Bonn. From 2009 to 2013 he did his Ph.D. research
in theoretical condensed matter physics in the group of Johann Kroha at the
University of Bonn. Since 2013 he has been working at the German Aerospace
Center focusing on quantum computing. In 2016 and 2017 he was a visiting
researcher at the Quantum Artificial Intelligence Laboratory at NASA Ames.

\end{IEEEbiography}
\begin{IEEEbiography}[{\includegraphics[width=1in,height=1.25in,clip,keepaspectratio]{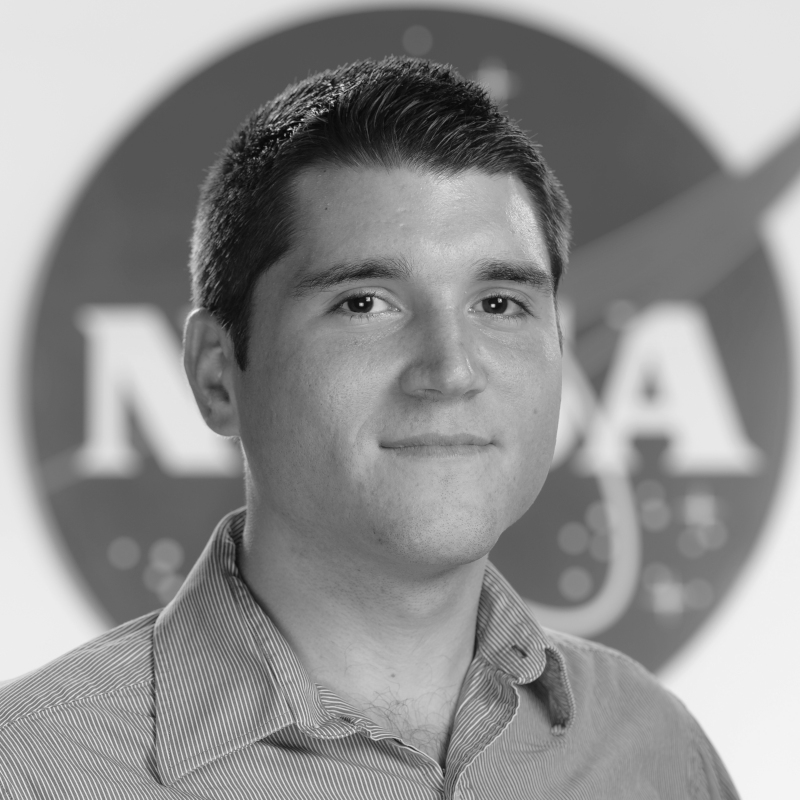}}]{Bryan O'Gorman}
received the A.B. degree in physics from Harvard College in 2013.
Since 2013 he has worked in the Quantum Artificial Intelligence Laboratory at
NASA Ames Research Center. He is currently pursuing his Ph.D. in computer
science at the University of California, Berkeley, as a NASA Space Technology
Fellow.

\end{IEEEbiography}
\begin{IEEEbiography}[{\includegraphics[width=1in,height=1.25in,clip,keepaspectratio]{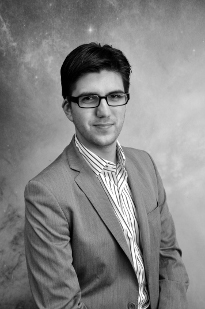}}]{Davide Venturelli}
graduated from Ecole Normale Superieure de Lyon and obtained his
Ph.D. in Numerical Simulations of the Condensed Matter at the International
School for Advanced Studies (SISSA) in Trieste and in Nanophysics at the
Universite de Grenoble (CNRS/UJF). In 2010-2011 he worked as a post-doc at
Scuola Normale Superiore in Pisa, Italy, in the Condensed Matter and Information
Theory group. 
%His early publications include quantum condensed matter many-body
%theory, design of nanodevices, quantum thermodynamics, quantum phase
%transitions, and non-equilibrium transport in mesoscopic systems. 
In 2012, he
joins the NASA Intelligent System Division (TI) as one of the founding members
of the Quantum Artificial Intelligence Laboratory (QuAIL), where he currently
works under the NASA Academic Mission Service contract. He is also Quantum
Computing team lead and Science Operations Manager of the Research Institute of
Advanced Computer Science (RIACS) at the Universities Space Research Association
(USRA).
%- one of the most prominent institutions providing workforce to NASA.
Venturelli is currently invested in research projects dealing with quantum
optimization applications and their near-term implementation in real hardware.
%His applied focus on algorithms is in advanced scheduling, telecommunication
%networks, robotics, AI planning, in collaborations with other governmental
%institutions, universities and the private sector.

\end{IEEEbiography}
\begin{IEEEbiography}[{\includegraphics[width=1in,height=1.25in,clip,keepaspectratio]{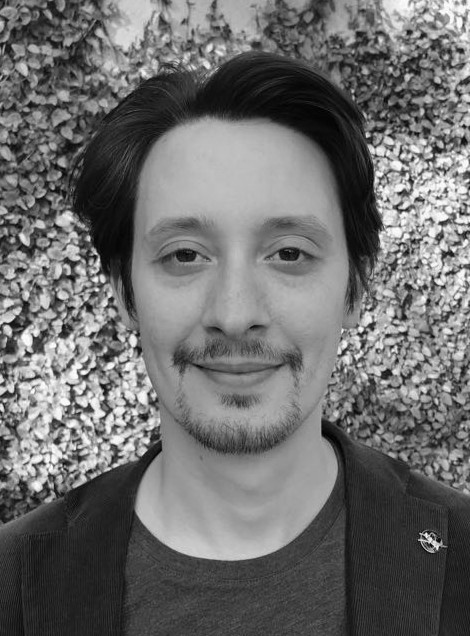}}]{Salvatore Mandr\`a}
obtained his Ph.D. in theoretical physics at the University of
Milan (Italy) in 2013. After the Ph.D, he worked as a postdoctoral researcher at
Harvard University and focused on quantum annealing and quantum computation. In
2016, he joined the Quantum Artificial Intelligence Lab (QuAIL) at NASA Ames.
His expertise ranges from the theoretical development of new classical/quantum
algorithms, as well as the numerical optimization of classical/quantum
simulations (including high level programming in C/C++ and distributed
programming in MPI/OpenMP).

\end{IEEEbiography}
\begin{IEEEbiography}[{\includegraphics[width=1in,height=1.25in,clip,keepaspectratio]{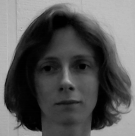}}]{Olga Rodionova}
received her Master's degree in Applied Mathematics and Computer
Science in 2011 from Saint-Petersburg State Polytechnical University
(Saint-Petersburg, Russia). She received her PhD in Applied Mathematics in 2015
from French Civil Aviation University (Toulouse, France). She then continued her
research in Air Traffic Management as a postdoctoral fellow at NASA Ames
Research Center (Moffett Field, CA, USA), working on aircraft trajectory and
airspace load optimization from 2015 to 2017. She is currently working as R\&D
engineer at Innov'ATM (Cugnaux, France).

\end{IEEEbiography}
\begin{IEEEbiography}[{\includegraphics[width=1in,height=1.25in,clip,keepaspectratio]{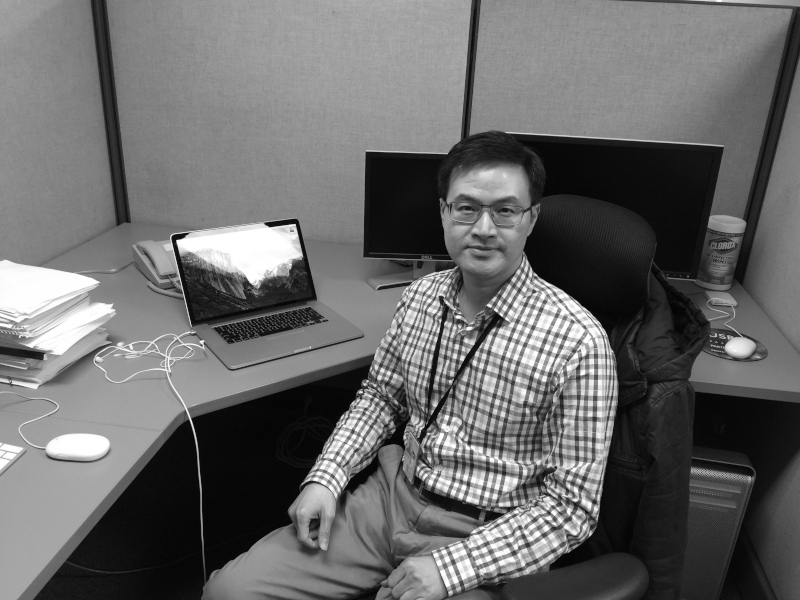}}]{Hok K. Ng}
is a research aerospace engineer at NASA Ames Research Center. He
specializes in algorithm development for air traffic management system. He has
worked for the past 10 years on Aviation related researches and applications. He
earned his Ph.D. degree in Mechanical and Aerospace Engineering from University
of California, Los Angeles. His current research interests include air traffic
management and disruptions management for Urban Air Mobility and Unmanned
Aircraft Systems.

\end{IEEEbiography}
\begin{IEEEbiography}[{\includegraphics[width=1in,height=1.25in,clip,keepaspectratio]{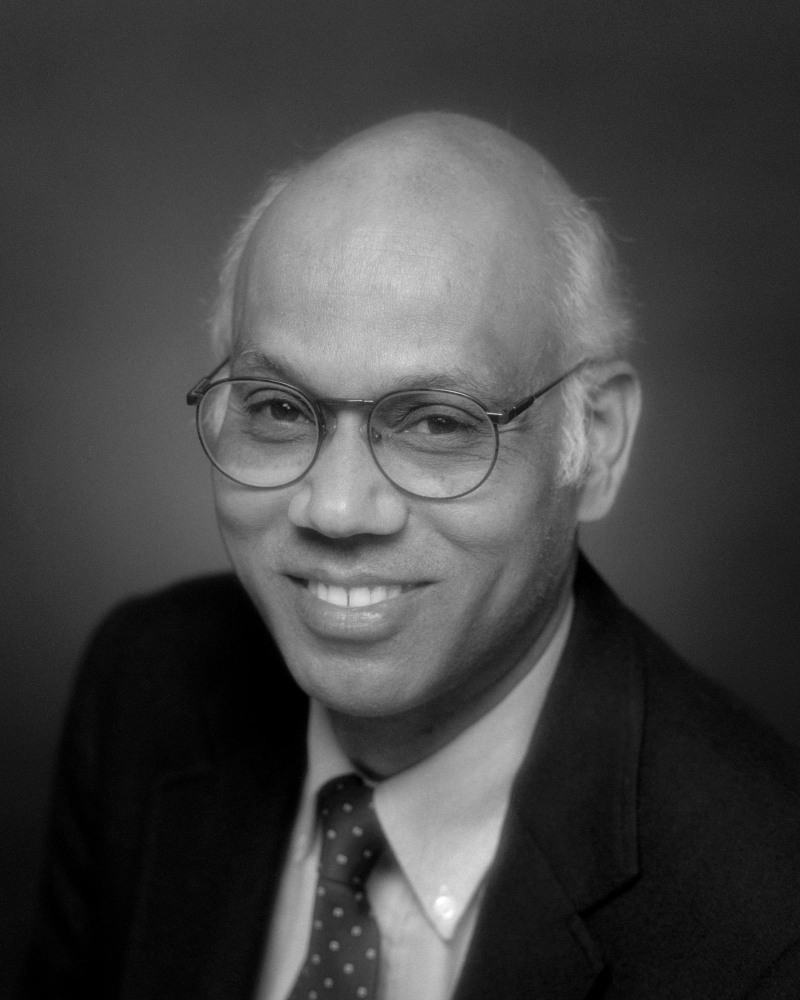}}]{Banavar Sridhar}
is a Research Associate at the NASA Ames Research Center. Earlier he served as
the NASA Senior Scientist for Air Transportation Systems. His research interests
are in the application of modeling and optimization techniques to aerospace
systems. Dr. Sridhar received the 2004 IEEE Control System Technology Award for
his contributions to the development of modeling and simulation techniques for
multi-vehicle traffic networks. He led the development of traffic flow
management software, Future ATM Concepts Evaluation Tool (FACET), which received
the AIAA Engineering Software Award in 2009, the NASA Invention of the Year
Award in 2010 and the FAA Award for the Excellence in Aviation Research in 2010.
He is a Fellow of the IEEE and the AIAA.

\end{IEEEbiography}
\begin{IEEEbiography}[{\includegraphics[width=1in,height=1.25in,clip,keepaspectratio]{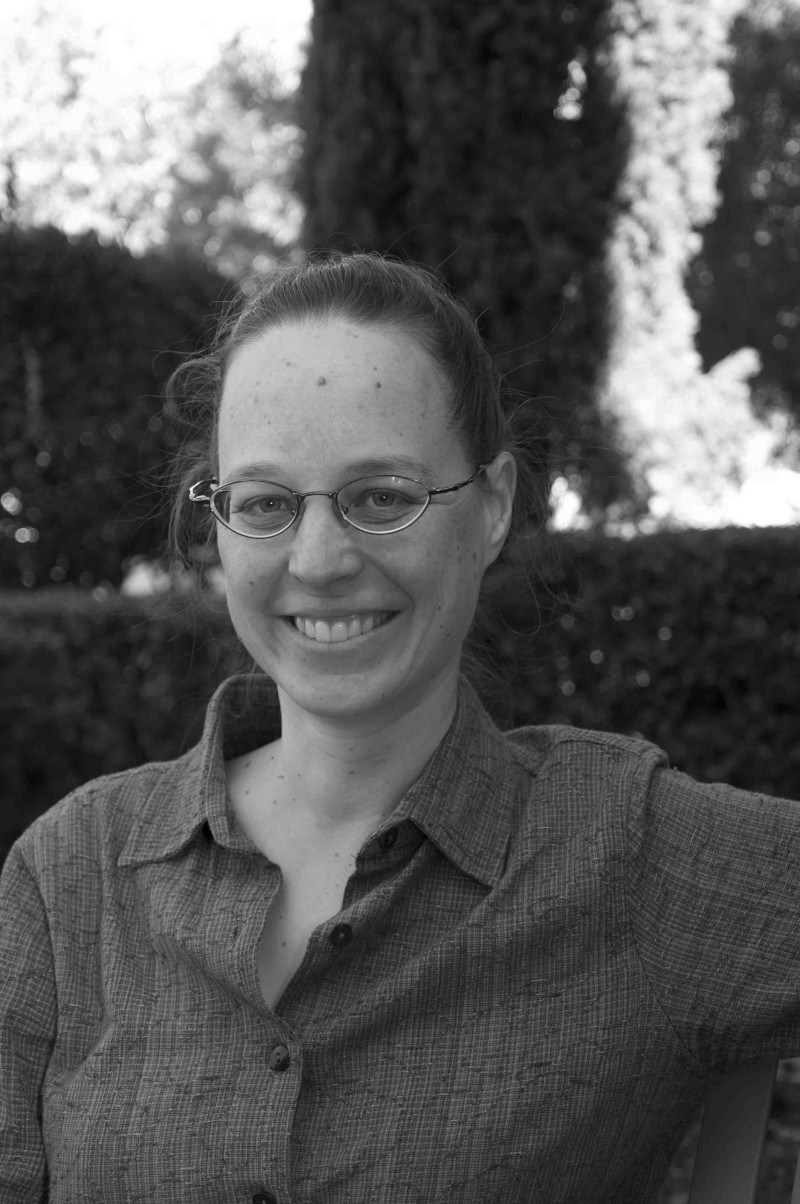}}]{Eleanor G. Rieffel}
leads the Quantum Artificial Intelligence Laboratory at the
NASA Ames Research Center. She joined NASA Ames Research Center in 2012 to work
on their expanding quantum computing effort, after working at FXPAL where she
performed research in diverse fields including quantum computation, applied
cryptography, image-based geometric reconstruction of 3D scenes, bioinformatics,
video surveillance, and automated control code generation for modular robotics.
Her research interests include quantum heuristics, evaluation and utilization of
near-term quantum hardware, fundamental resources for quantum computation,
quantum error suppression, and applications for quantum computing.  She received
her Ph.D. in mathematics from the University of California, Los Angeles.  She is
best known for her 2011 book Quantum Computing: A Gentle Introduction with
coauthor Wolfgang Polak and published by MIT press.

\end{IEEEbiography}
\begin{IEEEbiography}[{\includegraphics[width=1in,height=1.25in,clip,keepaspectratio]{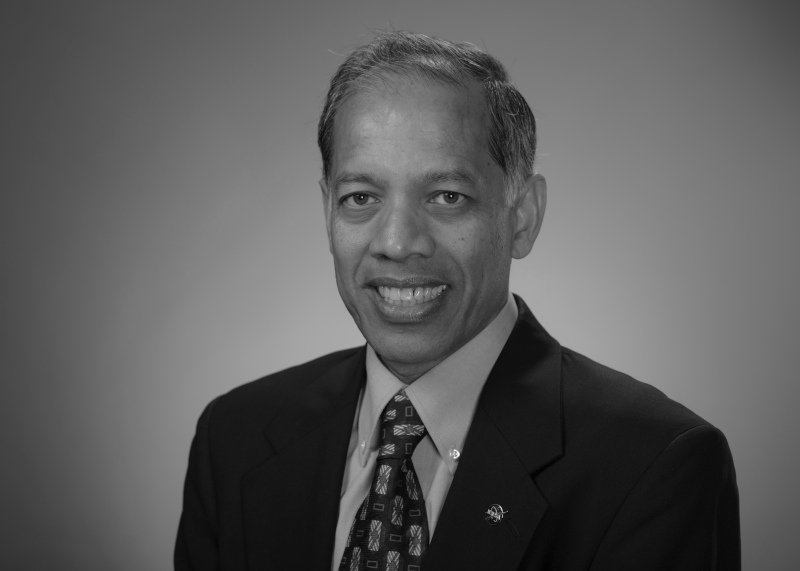}}]{Rupak Biswas}
is currently the Director of Exploration Technology at NASA
Ames Research Center, Moffett Field, California. In this role, he is in charge
of planning, directing, and coordinating the technology development and
operational activities of the organization that comprises of advanced
supercomputing, human systems integration, intelligent systems, and entry
systems technology. He is also the Manager of NASA’s High End Computing Project
that provides a full range of advanced computational resources and services to
numerous agency programs. Dr. Biswas received his Ph.D. in Computer Science from
Rensselaer in 1991, and has been at NASA ever since. He is an internationally
recognized expert in high performance computing and has published more than 160
technical papers, received many Best Paper awards, edited several journal
special issues, and given numerous lectures around the world.

\end{IEEEbiography}

\end{document}